%% file: acl_latex.tex
\documentclass[11pt]{article}
\usepackage[final]{acl}
\usepackage{pdflscape}
\usepackage{booktabs}
\usepackage{multirow}
\usepackage{longtable}
\usepackage{times}
\usepackage{latexsym}
\usepackage{amssymb}
\usepackage[T1]{fontenc}
\usepackage[utf8]{inputenc}
\usepackage{inconsolata}
\usepackage{graphicx}
\usepackage{kotex}
\usepackage{makecell}
\usepackage{booktabs}
\usepackage{siunitx}  
\usepackage{comment} 
\usepackage{listings}
\usepackage{tabularx} 
\usepackage{xcolor} 
\usepackage{enumitem}
\usepackage{multirow}
\usepackage[table]{xcolor}
\usepackage{algorithmic}
\usepackage{algorithm}
\usepackage{amsmath}

\usepackage[dvipsnames]{xcolor}

\title{Stride-$k$ Subsampling: Train-Free Audio Token Reduction for Whisper}

\author{
Chanhee Cho\thanks{\;\;Equal contribution.}  \quad
Junhyuk Choi\footnotemark[1] \quad
Bugeun Kim \\
\\
Department of Artificial Intelligence\\Chung-Ang University, Seoul, Republic of Korea \\ 
\texttt{\{cch991112, chwlnsgur129, bgnkim\}@cau.ac.kr}
}

\begin{document}
\maketitle
\begin{abstract}
Whisper exposes speech through a fixed 1500-token encoder interface, now a default representation for ASR decoders and Whisper-based speech language models (SpeechLMs), yet its redundancy remains largely unexamined. We propose stride-$k$ subsampling, a deterministic indexing operation that retains every $k$-th token after the convolutional stem or encoder transformer. Across five Whisper scales, $k=2$ preserves baseline WER at both positions, with CKA attributing this stability to acoustic overlap at the stem and attention-induced redistribution at the encoder output. Applying stride-2 at both positions cuts audio tokens by 75\% and total GFLOPs by 52--58\%, with small WER costs on most ASR benchmarks and larger costs on harder ones. The same configuration extends to three Whisper-based SpeechLMs, yielding modest accuracy drops on stronger baselines and larger drops on weaker ones, while reducing end-to-end latency by 19.6--27.4\%. Requiring no training or auxiliary computation, stride-$k$ subsampling exploits Whisper's preprocessing redundancy, indicating that its audio-token interface carries more capacity than downstream tasks require.

\end{abstract}

\section{Introduction}
Whisper \citep{radford2023robust} is widely used as the audio encoder for automatic speech recognition (ASR) and Whisper-based speech language models (SpeechLMs; \citet{ghosh2026audio,fang2025llama}). In these systems, the encoder output serves as the audio-token interface to the text-generating module, and for Whisper the interface is a fixed sequence of 1500 tokens. The token count governs downstream cost. ASR decoders cross-attend to the sequence at each autoregressive step \citep{monteiro2024xc,zelasko2025training}, and SpeechLMs insert the tokens into an LLM context \citep{lu2024fastadasp}. Reducing the number of encoder tokens is therefore a shared lever for lowering inference cost across both settings.

Whisper's preprocessing pipeline leaves the encoder input redundant. Whisper extracts log-mel features with a 25ms Hann window \citep{harris1978use} and a 10ms hop, and the convolutional stem aggregates several overlapping frames into each encoder token, giving every token an effective receptive field wider than its spacing. A calculation from the window and stride configuration shows that stride-2 subsampling after the stem introduces no temporal coverage gap. The retained tokens still cover the full input span, and their accumulated receptive-field overlap amounts to approximately 62\% of the span before any self-attention is applied.\footnote{See Appendix~\ref{app:overlap} for the derivation.} A substantial fraction of the tokens can therefore be dropped without retraining.

However, existing approaches to audio-encoder efficiency leave the redundancy at the encoder input unaddressed. Redesign-based methods rely on distillation \citep{gandhi2023distil, ferraz2024multilingual, waheed2024distill}, fine-tuning \citep{liu2024parameter, wang2025low}, or retraining \citep{sy2026baldwhisper, zhang2026whisper, orhon2025whisperkit}, and cannot be applied to deployed checkpoints without additional training. Train-free pruning methods avoid the retraining cost but share two limitations. First, they intervene on representations the encoder transformer has already processed---encoder hidden states \citep{xu2025early, lee2025token}, encoder outputs \citep{bhati2025towards}, or tokens passed to a downstream LLM \citep{lin2025speechprune, gibier2026segmentwise, jung2026fastav}---so the encoder input length is never reduced. Second, they select tokens through data-dependent criteria such as attention scores or pairwise similarity, which require auxiliary computation and produce input-specific token sets. Concurrently, \citet{xiang2026we} analyze deep-layer redundancy in LSLMs and propose Affinity Pooling, an adaptive cosine-similarity-based token merging method, and \citet{ltbm} propose LTBM, a train-free encoder-space token merging method that exploits locality.

Motivated by the preprocessing redundancy, we apply \textbf{stride-$k$ subsampling}, an operation that retains every $k$-th token and reduces sequence length through indexing alone. It adds no parameters and no auxiliary computation, and can be inserted into any Whisper-based model with a single indexing step. We apply it at two positions: immediately after the convolutional stem (\textit{input-side}), shortening the sequence fed into the encoder transformer, and after the encoder transformer (\textit{output-side}), shortening the sequence passed to the decoder or downstream LLM. Varying $k$ across five Whisper scales, we find that the two positions behave differently. \textit{Input-side} subsampling preserves WER up to $k=2$, whereas \textit{output-side} subsampling remains stable through $k=3$ and beyond---a gap the preprocessing overlap does not account for.

To examine the asymmetry, we measure adjacent-token similarity at the two positions with Centered Kernel Alignment (CKA; \citet{kornblith2019similarity}). At the \textit{input-side}, similarity decreases continuously as frame distance grows, consistent with the temporal overlap between adjacent frames. Stride-2 retains substantial accumulated overlap, whereas stride-3 reduces the overlap to only about 8\%, sharply weakening the redundancy buffer available after subsampling. This helps explain the abrupt degradation once $k$ exceeds 2. At the \textit{output-side}, similarity drops sharply between adjacent frames and then stays roughly flat, a pattern that may underlie the wider stable range there. The two positions therefore expose redundancy of different origins, and since the two sources are complementary, stride-2 can be applied at both positions simultaneously.

Compound stride-2 reduces the audio token count by 75\% and total GFLOPs by roughly 52--58\%, with WER largely retained across five Whisper scales on most ASR benchmarks and larger increases on more challenging ones. The same configuration extends without modification to three Whisper-based SpeechLMs, where task accuracy is largely retained on stronger baselines, and end-to-end latency improves by 19.6--27.4\%. Stride-$k$ subsampling requires no training or auxiliary computation, and serves as a drop-in efficiency improvement for Whisper-based models.

\section{Related Work}
Prior studies compress or redesign Whisper to reduce inference cost. Distillation-based approaches, such as Distil-Whisper \citep{gandhi2023distil}, compress the decoder depth using knowledge distillation, while others extend this framework to multilingual settings \citep{ferraz2024multilingual} or explore the robustness of the distillation process itself \citep{waheed2024distill}. Parameter-efficient fine-tuning methods \citep{liu2024parameter, wang2025low} adapt Whisper for low-resource environments by optimizing a small subset of parameters using techniques like LoRA \citep{hu2022lora}. Intervening at the hidden representation level offers another route. BaldWhisper \citep{sy2026baldwhisper} combines head shearing and layer merging, Whisper-MLA \citep{zhang2026whisper} converts the decoder's multi-head attention to multi-head latent attention to reduce key-value cache size, and WhisperKit \citep{orhon2025whisperkit} applies system-level optimizations and quantization. These approaches commonly require additional training procedures cannot be applied directly to already-deployed checkpoints.

A second line of work reduces audio token sequences without retraining. SpeechPrune \citep{lin2025speechprune} uses speech-text similarity and first-layer attention scores to remove audio tokens fed into a speech LLM, while Segmentwise Pruning \citep{gibier2026segmentwise} groups adjacent tokens into segments and prunes them via attention scores in audio-language models. Early Attentive Sparsification \citep{xu2025early} sparsifies hidden states using self-attention scores in early Whisper encoder layers, and Towards Audio Token Compression \citep{bhati2025towards} compresses encoder outputs through unsupervised segmentation and average pooling, aligning the result to a downstream LLM via low-rank adapters. Concurrently, \citet{xiang2026we} analyze deep-layer redundancy in LSLMs and propose Affinity Pooling, an adaptive cosine-similarity-based token merging method, and \citet{ltbm} propose LTBM, which merges encoder-space audio tokens by pairwise similarity under an explicit temporal-locality constraint, so that a token is combined only with neighbors inside a fixed time window.

These studies show that audio-sequence redundancy is a useful efficiency signal, but two gaps remain. First, prior methods target hidden states downstream of the audio encoder, tokens passed to a downstream LLM, or intermediate encoder tokens, and therefore do not reduce input length before the encoder transformer processes it. Second, they rely on data-dependent criteria such as similarity computation or attention score analysis.


\section{Stride-$k$ Subsampling}
\label{sec3}
\subsection{Preliminaries}
Whisper extracts log-mel features with a 25\,ms Hann window and 10\,ms hop, and a convolutional stem (conv stem) with stride-2 produces a 1500-token sequence. The encoder transformer preserves sequence length, and the decoder cross-attends to all 1500 tokens at every autoregressive step, making decoder cross-attention FLOPs linear in encoder output length. The 25 ms window and 10 ms hop introduce overlap between adjacent log-mel frames, and the convolutional stem further expands this redundancy by aggregating multiple overlapping frames into each encoder token. An analysis that accounts for the STFT window, hop size, and the effective receptive field induced by the convolutional stem shows that stride-2 subsampling after the stem introduces no temporal coverage gap while still leaving approximately 62\% of the input span overlapped by adjacent retained receptive fields. At stride-3, this accumulated overlap drops to only 8.3\%, sharply weakening the redundancy buffer available after subsampling.\footnote{Refer to Appendix~\ref{app:overlap}} We define stride-$k$ subsampling as $H’$=\texttt{H[::k,:]}$ \in \mathbb{R}^{\lceil T/k \rceil \times d}$ for an encoder representation $H \in \mathbb{R}^{T \times d}$ with $T$ tokens and hidden size $d$. The operation requires no training, fine-tuning, distillation, or attention-based selection, and reduces sequence length to $1/k$ via indexing alone. We apply it at two positions (\textit{input-side} and \textit{output-side}). For the full algorithm and implementation details, refer to Appendix~\ref{app:stridekdetail}.

\subsection{Input-side Stride-$k$ Subsampling}
\textit{Input-side} subsampling applies stride-$k$ subsampling immediately after the conv stem, before the encoder transformer. The encoder processes a sequence of length 1500/$k$ instead of 1500, reducing both encoder self-attention FLOPs and decoder cross-attention FLOPs. \textit{Input-side} subsampling directly exploits the acoustic redundancy induced by the window hop length and conv stem.

\subsection{Output-side Stride-$k$ Subsampling}
\textit{Output-side} subsampling applies stride-$k$ subsampling immediately after the encoder transformer, before decoder cross-attention. In the vanilla encoder path, the encoder still processes all 1500 tokens, so encoder FLOPs are unchanged; only the decoder cross-attention FLOPs are reduced through the smaller number of target tokens. \textit{Output-side} subsampling exploits the representational redundancy formed between adjacent tokens by encoder self-attention.

\section{Diagnosing the stride-$k$ Subsampling}
\label{sec4}
\subsection{Experimental Setup}
\subsubsection{Benchmark}
To analyze the effect of stride-$k$ subsampling, we use three benchmarks spanning distinct acoustic conditions and a range of recognition difficulty. LibriTTS \citep{zen2019libritts} consists of clean read speech. ESD \citep{zhou2022emotional} consists of emotionally expressive speech recorded in a controlled environment. Common Voice \citep{ardila2020common} consists of crowd-sourced speech under varied noise and recording conditions.

\begin{figure*}[t]
    \centering
    \includegraphics[width=\textwidth]{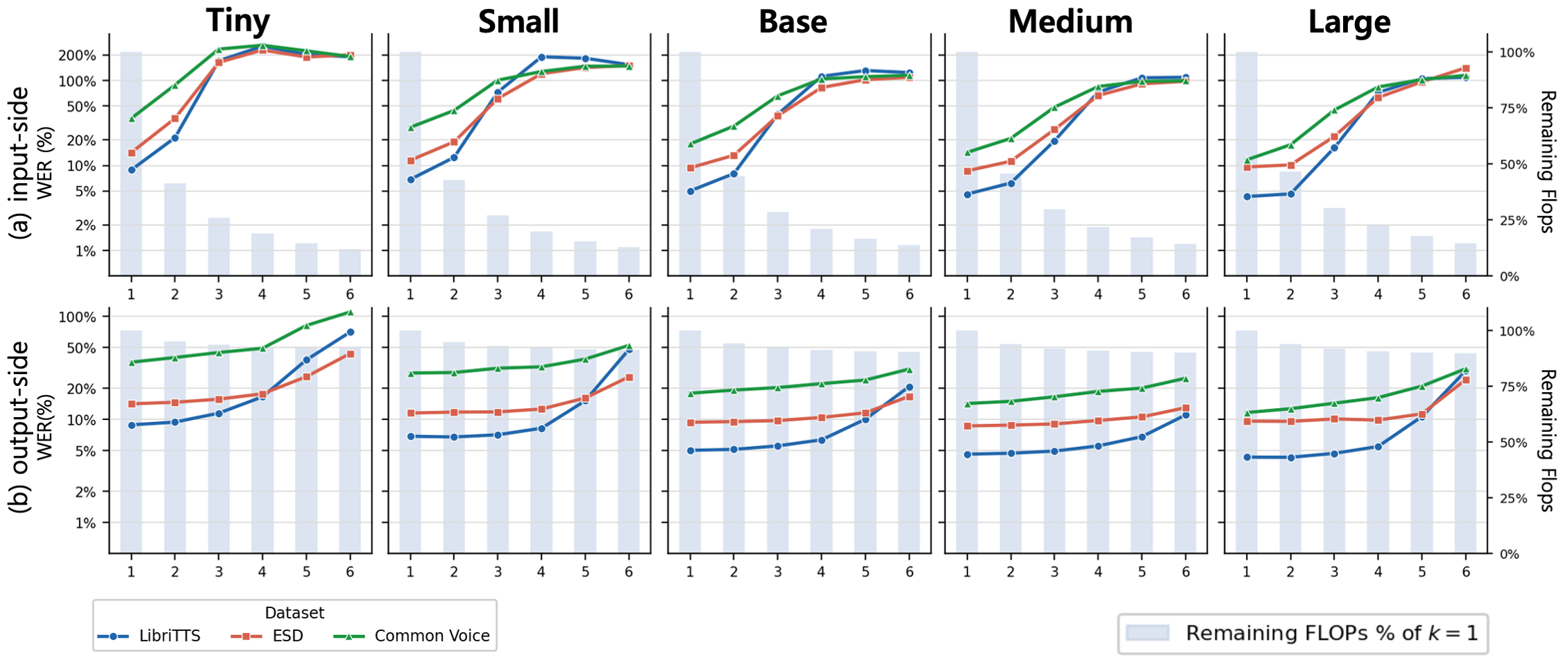} 
    \caption{WER under stride-$k$ subsampling. Top row: \textit{input-side}; bottom row: \textit{output-side}. Left $y$-axis: WER (lines, one per benchmark); right $y$-axis: remaining FLOPs relative to $k{=}1$ (bars).}
    \label{fig:eachexperimentresult}
\end{figure*}

\subsubsection{Selected Models}
We evaluate stride-$k$ subsampling on all five Whisper scales (tiny, base, small, medium, large) for $k\in\{1, \cdots, 6\}$, where $k=1$ denotes the unmodified baseline. We focus on Whisper because it is one of the most widely used encoder--decoder ASR models and is also commonly used as a speech encoder in recent SpeechLMs. For Whisper, the STFT window, hop size, and convolutional-stem stride imply that retained post-stem tokens continue to overlap up to $k=3$; evaluating through $k=6$ allows us to examine behavior beyond this overlap regime. Detailed settings are in Appendix~\ref{app:diagonisesetup}.

\subsubsection{Evaluation Metrics}
We measure both ASR performance and inference cost. ASR performance is reported as Word Error Rate (WER). Inference cost is measured by the total FLOPs. Both quantities scale linearly with the number of encoder output tokens, and are therefore reduced by a factor of 1/$k$ under stride-$k$.

\subsection{Results}
We report WER across three benchmarks, five Whisper scales, and stride values $k{=}1,\cdots,6$ at both subsampling positions, analyzed along three axes: benchmark, model scale, and subsampling position. Figure~\ref{fig:eachexperimentresult} summarizes the results, with full per-configuration numbers in Appendix~\ref{app:detailResultDiagonise}.

\paragraph{Stability scales with input difficulty.}
The stable range of $k$ correlates with baseline WER. On LibriTTS, baseline WER for Whisper-Large is 4.31 and \textit{output-side} WER is 4.30 at $k{=}2$ and 5.47 at $k{=}4$. On Common Voice, baseline WER for Whisper-Large is 11.66 and \textit{output-side} WER rises to 12.70 at $k{=}2$ and 16.26 at $k{=}4$. The \textit{input-side} position shows the same ordering: at $k{=}2$, the WER increase relative to baseline is smallest on LibriTTS and largest on Common Voice. ESD shows higher absolute WER than LibriTTS but a comparable relative trajectory across $k$.

\paragraph{Stability scales with model scale.}
Within each benchmark, larger models tolerate larger $k$ before degradation. On LibriTTS at \textit{input-side} $k{=}2$, WER relative to baseline increases by a factor of 2.40 for Tiny (8.86 $\rightarrow$ 21.23), 1.82 for Base (6.87 $\rightarrow$ 12.47), 1.60 for Small (5.02 $\rightarrow$ 8.04), 1.35 for Medium (4.61 $\rightarrow$ 6.22), and 1.07 for Large (4.31 $\rightarrow$ 4.63). At output-side $k{=}4$ on LibriTTS, WER stays within one point of baseline for Medium (4.61 $\rightarrow$ 5.54) and Large (4.31 $\rightarrow$ 5.47), while Tiny rises from 8.86 to 16.57 and Base from 6.87 to 8.22. The same monotonic ordering holds on ESD; on Common Voice the ordering holds but the absolute gap between scales is larger.

\paragraph{Input-side and output-side degrade differently.}
The two positions differ in both the stable range of $k$ and the shape of degradation beyond it. At $k{=}2$, both remain close to baseline, though the \textit{output-side} WER is consistently lower than the \textit{input-side} WER at the same $k$ (LibriTTS Whisper-Small: input 8.04 vs.\ output 5.13; LibriTTS Whisper-Medium: input 6.22 vs.\ output 4.70). At $k{=}3$, \textit{input-side} WER increases abruptly across all benchmarks and scales (LibriTTS Whisper-Medium: 4.61 $\rightarrow$ 19.51; Whisper-Large: 4.31 $\rightarrow$ 16.20), while output-side WER remains within a small margin of baseline (Whisper-Medium: 4.61 $\rightarrow$ 4.94; Whisper-Large: 4.31 $\rightarrow$ 4.69). The asymmetry persists at $k{=}4$, with \textit{input-side} WER climbing steeply and \textit{output-side} WER increasing gradually. The pattern is consistent across the benchmark and scale axes reported above.

\subsection{Discussion}
\paragraph{Stability tracks baseline strength.}
The benchmark and scale results share a common pattern. Configurations with lower baseline WER tolerate larger $k$ before degradation. The pattern holds across both subsampling positions, indicating that stride-$k$ subsampling is most reliably applicable when the underlying Whisper checkpoint already performs well on the input. The behavior is consistent with the interpretation that subsampling exposes a fixed fraction of redundancy, and that recognition under reduced input is bounded by the model's baseline competence on that input.

\paragraph{Preprocessing overlap does not explain the output-side.}
Output-side subsampling stays near baseline over a wider $k$ range than input-side, degrading gradually rather than abruptly. The Section~\ref{sec3} overlap argument explains \textit{input-side} $k{=}2$ via the receptive field and window overlap between adjacent frames, but not the wider \textit{output-side} range, where adjacent tokens come from the encoder transformer rather than the conv stem. This suggests the encoder output carries redundancy absent after the conv stem, which we probe in Section~\ref{sec:cka} via adjacent-token similarity.

\section{CKA Analysis of Adjacent-token Similarity}
\label{sec:cka}
The positional asymmetry in Section~\ref{sec4} is not predicted by the frame-overlap argument alone, which accounts only for the local redundancy at the conv stem. To investigate the wider stable range at the \textit{output-side}, we measure how representational similarity between tokens evolves with frame distance at the two positions.

\subsection{Experimental Setup}

\paragraph{Data.}
We run the analysis on LibriTTS. For each utterance in the test set, we extract token representations at the conv stem output and the encoder output, and select $10$ anchor tokens from positions with a full neighborhood up to ${d=}5$. Measurements are performed across the five Whisper scales. Further details are in Appendix~\ref{app:ckasetupDetail}.

\paragraph{Measurement.}
For each position and each frame distance $d \in \{1,\cdots,5\}$, we compute Linear CKA between the anchor tokens and the tokens at offset $d$. Linear CKA measures the alignment of representational geometry between two sets of vectors and is invariant to orthogonal transformation and isotropic scaling. Absolute CKA values, however, are not directly comparable across the two positions, since conv stem outputs and encoder outputs lie in different representation spaces with different intrinsic self-similarity scales (see Appendix~\ref{app:ckasetupDetail}). We therefore report, as our primary quantity, the step-wise change

{\small
\[
\Delta(d{\to}d{+}1)
=
100 \times
\bigl(
\text{CKA}(d{+}1)-\text{CKA}(d)
\bigr)
/\text{CKA}(d).
\]
}
which characterizes the shape of similarity decay independently of its absolute level.


\begin{table}[t]
\centering
\small
\begin{tabular}{@{}l@{\;\;}l@{\;\;}r@{\;\;}r@{\;\;}r@{\;\;}r@{\;\;}r}
\toprule
\textbf{Scale} & \textbf{Layer} & $d=0$ & $d=1$ & $d=2$ & $d=3$ & $d=4$ \\
\midrule
\multirow{2}{*}{Tiny}
& Conv Stem & -3.60 & -6.26 & -5.82 & -4.63 & -2.99 \\
& Enc Out   & -11.92 & -11.96 & -4.87 & -0.24 & +0.37 \\
\midrule
\multirow{2}{*}{Base}
& Conv Stem & -3.59 & -6.48 & -6.21 & -5.05 & -3.25 \\
& Enc Out   & -9.70 & -10.66 & -5.44 & -1.85 & +0.55 \\
\midrule
\multirow{2}{*}{Small}
& Conv Stem & -3.92 & -6.90 & -6.57 & -5.28 & -3.37 \\
& Enc Out   & -17.23 & -9.80 & +0.99 & -1.43 & +3.16 \\
\midrule
\multirow{2}{*}{Medium}
& Conv Stem & -3.81 & -6.20 & -5.62 & -4.53 & -3.04 \\
& Enc Out   & -21.33 & -15.82 & -2.11 & +0.35 & +0.37 \\
\midrule
\multirow{2}{*}{Large}
& Conv Stem & -2.05 & -3.01 & -2.63 & -2.14 & -1.39 \\
& Enc Out   & -14.03 & -11.26 & -5.07 & -1.50 & -0.49 \\
\bottomrule
\end{tabular}
\caption{Step-wise change
$\Delta(d{\to}d{+}1)=100\cdot
(\mathrm{CKA}(d{+}1)-\mathrm{CKA}(d))/\mathrm{CKA}(d)$.
{\footnotesize where $\mathrm{CKA}(0)=1$}}
\label{tab:cka_step}
\vspace{-1em}
\end{table}

\subsection{Results and Discussion}
\paragraph{Conv stem decays continuously while encoder output plateaus after $d{=}1$.}
Table~\ref{tab:cka_step} shows two distinct decay shapes. At the conv stem, $\Delta$ stays negative and shrinks monotonically in magnitude as $d$ grows, a continuous decay: on Whisper-Base it moves from $-6.48$ at $d{=}1$ to $-6.21$, $-5.05$, $-3.25$ at $d{=}2,3,4$, and the same profile holds across scales. The smaller magnitude at $d{=}0$ reflects the trivial self-similarity reference $\text{CKA}(0){=}1$, so we read the decay shape from $d{=}1$ onward. This continuous profile follows from receptive-field overlap: each post-stem token spans an effective receptive field of about $65$\,ms while adjacent tokens are spaced every $20$\,ms, and the resulting shared acoustic support diminishes gradually as distance grows. At the encoder output, the relative change is large over the first two steps and then becomes much smaller: on Whisper-Large it moves from $-14.03$ and $-11.26$ at $d{=}0,1$ to $-5.07$, $-1.50$, $-0.49$ at $d{=}2,3,4$, with the same pattern across scales. Beyond $d{=}1$, similarity between an anchor and its neighbors no longer depends systematically on distance. This suggests that the encoder transformer, while still locally biased, does not concentrate redundancy only in the adjacent pair; instead, it redistributes redundancy across a wider local neighborhood.

\paragraph{The two decay shapes align with the WER.}
The two shapes correspond to the two degradation patterns observed under stride-$k$ subsampling. At the \textit{input-side}, similarity decays continuously, so the retained tokens become progressively weaker substitutes for the dropped ones as $d$ grows. This matches the \textit{input-side} WER profile, which stays near baseline at $d{=}2$ and rises sharply from $d{=}3$ onward. At the \textit{output-side}, the similarity change is concentrated in the first two steps and plateaus afterward, so additional skipping incurs little further loss, matching the gradual \textit{output-side} WER profile through $d{=}4$. The positional asymmetry in Section~\ref{sec4} therefore reflects a difference in the shape of similarity decay, not in its absolute level.

\paragraph{The output-side plateau exceeds what locality predicts.}
The Whisper encoder is reported to attend locally, with self-attention dominated by a diagonal band~\citep{zhang2026speaking,alastruey2022locality}. A purely local account would predict a continuous decay at the encoder output, as at the conv stem, but the observed plateau departs from this. The encoder transformer, though still local, redistributes redundancy across the neighborhood rather than concentrating it in the adjacent pair. This redundancy is independently consistent with reports that the upper Whisper encoder layers can be pruned without loss~\citep{irigoyen2025pruning}, and is what supports the wider stable range of $d$ at the \textit{output-side}. Full CKA values are in Appendix~\ref{app:cka_full}.

\section{Compound Stride-2 
}
\label{sec:final}
\input{finalResult}
Two findings motivate combining the positions. First, stride-2 preserves baseline WER at both the \textit{input-side} and \textit{output-side} across all scales and benchmarks. Second, the stability arises from independent forms of overlap—one at the conv stem, one at the encoder output—that can be exploited together. We therefore apply stride-2 at both positions and evaluate it as a drop-in efficiency improvement for Whisper-based models.

\subsection{Experiment Setup}
We evaluate compound stride-2 ($k_{\mathrm{in}}=k_{\mathrm{out}}=2$) in two settings, Whisper ASR and SpeechLMs. Subsampling is applied only to the Whisper encoder, leaving all other components unmodified. Reproduction details are in Appendix~\ref{app:Compundsetup}.

\paragraph{Whisper ASR.}
Since Section~\ref{sec4} shows that stability under stride-$k$ improves with model scale, we focus on the three larger Whisper scales—Small, Medium, and Large-v3—and evaluate on LibriTTS, ESD, and English Common Voice, with Chinese Common Voice results reported in Appendix~\ref{app:multilingual}. Performance is measured by WER, and efficiency by total GFLOPs and end-to-end latency.

\paragraph{SpeechLMs.} We evaluate three widely used open-source models, Audio Flamingo~3 (AF3; \citealp{ghosh2026audio}), Qwen2-Audio \citep{chu2024qwen2}, and LLaMA-Omni~2 \citep{fang2025llama}, on two benchmarks, MMSU \citep{dingdong2026mmsu} and MMAU \citep{sakshi2025mmau}. MMSU evaluates spoken language understanding through linguistic and paralinguistic tasks, split into a perception group that probes fine-grained acoustic cues such as intonation, pause, and stress, and a reasoning group requiring higher-level inference over them. MMAU covers general audio reasoning across the sound, music, and speech domains, and requires expert-level knowledge beyond direct acoustic perception. Across both benchmarks, some categories---intonation and pause in MMSU, counting in MMAU---depend on fine-grained temporal structure, the timing of acoustic events rather than their identity, which we revisit in the failure-case analysis. Performance is measured by accuracy, and efficiency by total GFLOPs and end-to-end latency. To analyze where subsampling changes model behavior rather than only by how much, we additionally report per-example hurt and help rates, the fractions of baseline-correct examples that become incorrect and incorrect examples that become correct under compound stride-2.

\subsection{Results}
\paragraph{ASR.}
Compound stride-2 keeps WER reasonably close to baseline on LibriTTS and ESD, with the gap shrinking as model scale increases (Table~\ref{tab:compound-main}). On LibriTTS, the WER increase decreases from $+4.20$ on Small to $+2.40$ on Medium and $+1.07$ on Large-v3; on ESD, the corresponding gaps are $+4.84$, $+3.50$, and $+1.01$. Common Voice is the most sensitive benchmark, showing substantially larger increases of $+13.40$, $+9.56$, and $+9.90$. These results indicate that compound stride-2 is absorbed more reliably by stronger Whisper checkpoints and by cleaner or more controlled inputs. Although total GFLOPs drop by approximately $57$\%, end-to-end latency changes only modestly ($+3.6$\%, $-1.1$\%, and $-6.0$\%), which we analyze in Section~\ref{sec:final}.

\paragraph{Whisper-based SpeechLMs.}
Across the three models, task accuracy decreases with compound stride-2, and the size of the drop varies systematically. Audio Flamingo~3 shows the smallest drops on every block, with $-1.00$ on MMSU Perception, $-2.81$ on MMSU Reasoning, and $-1.7$ on MMAU Audio. Qwen2-Audio falls between AF3 and LLaMA-Omni~2 ($-1.62$, $-6.16$, and $-3.6$, respectively), and LLaMA-Omni~2 shows the largest drops ($-3.06$, $-9.71$, $-5.5$). The ordering is consistent across the three benchmark blocks. Within each model, MMSU Perception is more robust than MMSU Reasoning, with the largest individual drop occurring on LLaMA-Omni~2 Reasoning Linguistic Semantics ($-13.72$). Total GFLOPs drop by $51.9$--$56.6$\% across the three models, and end-to-end latency drops by $19.6$--$27.4$\%. The hurt and help rates show that these average drops come from asymmetric example-level changes. MMSU Perception keeps hurt and help close, as in AF3 Linguistic Semantics ($3.46$ vs. $3.15$), while MMSU Reasoning widens the gap, with hurt exceeding help by $5.69$, $8.75$, and $13.72$ points on Linguistic Semantics for AF3, Qwen2-Audio, and LLaMA-Omni~2. MMAU splits the same way, with Music staying balanced and LLaMA-Omni~2 Speech ($14.71$ vs. $4.50$) and Sound ($11.41$ vs. $4.20$) showing the largest hurt rates.

\subsection{Discussion}

\paragraph{Robustness follows baseline and input stability.}
The aggregate results show that compound stride-2 is absorbed more reliably when the underlying system already performs well. Across SpeechLMs, the accuracy-drop ordering follows the baseline ordering: AF3 is the most stable, Qwen2-Audio is intermediate, and LLaMA-Omni~2 is the most sensitive. The ASR results show the same trend across Whisper scales, where Large-v3 incurs smaller WER gaps than Small and Medium on LibriTTS and ESD. Input condition also matters. Common Voice shows the largest WER increase, likely because its noisier and more heterogeneous recordings leave less redundancy for temporal subsampling, whereas the cleaner LibriTTS and controlled ESD remain closer to baseline. A controlled white-noise sweep on LibriTTS confirms this directly. As SNR falls from clean to 0\,dB, the compound stride-2 gap grows from $+0.9$ to $+28.6$ WER on Large-v3 and from $+3.7$ to $+57.5$ on Small, while output-side $k=2$ stays within about one point of baseline at every level (Appendix~\ref{app:noise}). The same pattern reproduces beyond English. On Common Voice Chinese, measured in character error rate, both the scale ordering and the input/output asymmetry hold, with output-side $k=2$ staying near baseline at every scale (Appendix~\ref{app:multilingual}). Thus, compound stride-2 is most reliable when both the model and the input domain provide enough robustness margin under reduced temporal resolution. These aggregate drops, however, do not reveal which examples change. We examine this next through the hurt/help rates.

\paragraph{Error shifts depend on temporal evidence.}
The hurt and help rates reveal where, not just how much, compound stride-2 changes model behavior. Hurt cases concentrate in groups that require sparse or temporally localized evidence, with MMSU Reasoning more fragile than Perception and MMAU Sound and Speech more fragile than Music. These groups rely on cues such as pause boundaries, word-level intonation, and event counting, where the answer depends not on what acoustic event is present but on when it occurs, and compound stride-2 reduces the temporal resolution, blurring the few frames that determine the answer. Help cases show the opposite tendency, appearing when the decision can rely on broader patterns such as coarse prosody, tempo, or stable timbre. This suggests compound stride-2 does not always behave as pure information loss, and in some cases acts as a temporal smoothing operation that suppresses local distractors. The method is therefore most reliable when the downstream decision depends on broadly distributed acoustic evidence, and most fragile when it requires sparse, fine-grained temporal cues.

\paragraph{Transfer depends on the frontend, not the architecture.}
The conditions above concern when stride-2 is safe within a Whisper-based system; a separate question is which encoders admit it at all. We do not claim that stride-$k$ generalizes to every speech encoder. Appendix~\ref{app:beyondwhisper} evaluates this question on additional encoders, including the Phi-4-multimodal conformer and raw-waveform encoders such as wav2vec~2.0, HuBERT, and WavLM. The operative condition is narrower, and can be checked before running a model: whether the preprocessing pipeline leaves adjacent tokens overlapping enough that the retained ones still cover the signal, that is, $R \ge kS$ for receptive field $R$ and spacing $S$ (Appendix~\ref{app:overlap}). Whisper satisfies it ($R=65$\,ms at $S=20$\,ms), and so does the mel-plus-convolution frontend of the Phi-4-multimodal conformer ($165$\,ms at $80$\,ms), which tolerates compound stride-2 with larger but non-catastrophic drops. Raw-waveform encoders do not: wav2vec~2.0, HuBERT, and WavLM feed the waveform directly into convolutions ($25$\,ms at $20$\,ms) and collapse at the first stride ($3.06 \rightarrow 17.21$, $3.68 \rightarrow 24.55$, and $7.60 \rightarrow 48.67$ WER). Holding the encoder blocks fixed and varying only the frontend isolates the cause: a raw-waveform conformer collapses like the self-supervised encoders ($4.04 \rightarrow 28.10$), while the mel-input conformer does not. Since overlapping windowed analysis is standard practice in signal processing, any encoder following that convention is a candidate for stride-$k$.

\paragraph{Compute reduction translates differently to wall-clock gains across settings.}
Compound stride-2 reduces total GFLOPs by a similar amount in the two settings ($\sim$57\% for ASR, $52$--$57$\% for SpeechLMs), but end-to-end latency improves much more in SpeechLMs ($-19.6$\% to $-27.4$\%) than in ASR ($-6.0$\% on Large, near-zero or slightly negative on smaller scales). The asymmetry follows from where audio tokens enter the computation. In Whisper ASR, the decoder is autoregressive while the encoder runs once, and the decoder dominates total runtime~\citep{radford2023robust,kwon2025whisfusion}. Compound stride-2 reduces the number of cross-attention keys but leaves the number of decoder steps unchanged, so the FLOPs reduction does not transfer to speedup. In SpeechLMs, audio tokens are prepended to a downstream LLM and consumed during prefill, so audio token count determines prefill cost. Reducing audio tokens from 1500 to 375 lowers prefill cost in proportion to the token reduction, and end-to-end latency improves accordingly.

\subsection{Composability with other methods}
Stride-$k$ subsampling reduces tokens at the encoder, whereas existing train-free methods prune tokens after they enter the downstream LLM. To examine how the two interact, we compare against SpeechPrune (SP), a representative attention-based pruning method for speech LLMs, at a pruning ratio matched to the 75\% token reduction of compound stride-2 for an equal compression budget. Beyond this head-to-head comparison, the two methods can also be combined: applying \textit{input-side} stride-2 on top of SP reduces compute further with little additional accuracy loss, indicating that the two operate on complementary redundancy and compose rather than compete. Full results are in Appendix~\ref{app:spDetail}.

Stride-$k$ also composes with distillation. Distil-Whisper compresses the decoder, an axis orthogonal to the encoder-side token reduction of stride-$k$, so the two can be stacked. Applying compound stride-2 to Distil-Whisper large-v3 cuts its total FLOPs by roughly a further 54\%, with WER staying close to the Distil-Whisper baseline on LibriTTS and ESD and a larger increase on the harder Common Voice, consistent with the difficulty trend in Section~\ref{sec:final} (Appendix~\ref{app:distilwhisper}).

\begin{table}[t]
\centering
\small
\setlength{\tabcolsep}{4pt}
\begin{tabular}{l ccc}
\toprule
\multicolumn{4}{l}{\textit{(a) vs.\ SpeechPrune --- SpeechLMs, Accuracy}} \\
\cmidrule(l){1-4}
 & AF3 & Qwen2-A. & L-Omni2 \\
\midrule
\multicolumn{4}{l}{\textit{MMSU Perception}} \\
\cmidrule(l){2-4}
Base.          & 42.67 & 41.12 & 29.96 \\
SP             & 38.02 & 33.88 & \textbf{27.05} \\
\textbf{Ours}  & \textbf{41.67} & \textbf{39.50} & 26.90 \\
\midrule
\multicolumn{4}{l}{\textit{MMSU Reasoning}} \\
\cmidrule(l){2-4}
Base.          & 77.19 & 69.71 & 58.47 \\
SP             & 62.73 & 53.60 & \textbf{50.50} \\
\textbf{Ours}  & \textbf{74.38} & \textbf{63.55} & 48.76 \\
\midrule
\multicolumn{4}{l}{\textit{MMAU}} \\
\cmidrule(l){2-4}
Base.          & 70.5 & 56.3 & 51.4 \\
SP             & 66.1 & 49.0 & \textbf{47.0} \\
\textbf{Ours}  & \textbf{68.8} & \textbf{52.7} & 45.9 \\
\midrule
\multicolumn{4}{l}{\textit{Efficiency}} \\
\cmidrule(l){2-4}
SP $\Delta$FLOPs            & $-$33.13\% & $-$32.45\% & $-$39.91\% \\
\textbf{Ours} $\Delta$FLOPs & \textbf{$-$52.8\%} & \textbf{$-$51.9\%} & \textbf{$-$56.6\%} \\
SP $\Delta$Lat.             & $-$20.02 & $-$14.56 & $-$24.37 \\
\textbf{Ours} $\Delta$Lat.  & \textbf{$-$22.3\%} & \textbf{$-$19.6\%} & \textbf{$-$27.4\%} \\
\midrule
\multicolumn{4}{l}{\textit{(b) vs.\ Distil-Whisper --- ASR, WER}} \\
\cmidrule(l){1-4}
 & WER & $\Delta$WER & $\Delta$FLOPs \\
\midrule
LibriTTS       & 8.31  & $+3.37$  & $-$54.3\% \\
ESD            & 13.28 & $+3.80$  & $-$54.3\% \\
Common Voice   & 31.07 & $+16.56$ & $-$54.3\% \\
\bottomrule
\end{tabular}
\caption{Composability of stride-$k$. \textbf{(a)} vs.\ SpeechPrune (SP) on Whisper-based SpeechLMs; Efficiency rows are $\Delta$ vs.\ Baseline. \textbf{(b)} Compound stride-2 on top of Distil-Whisper large-v3; $\Delta$ vs.\ the Distil-Whisper.}
\label{tab:sp_compare}
\vspace{-2em}
\end{table}

\section{Conclusion}

We propose stride-$k$ subsampling, a train-free indexing operation that reduces Whisper audio tokens by retaining every $k$-th token after the convolutional stem or encoder transformer. Across five Whisper scales, $k=2$ keeps WER close to baseline, with CKA linking this stability to acoustic overlap and attention-induced redistribution. Compound stride-2 cuts audio tokens by 75\% and GFLOPs by 52--58\%, with small WER costs on most ASR benchmarks and larger costs on harder ones. On three Whisper-based SpeechLMs, it yields modest accuracy drops on stronger baselines and larger drops on weaker ones, while reducing latency by 19.6--27.4\%. The latency results show that audio-token reduction has its largest wall-clock effect when tokens enter downstream LLM prefill, while ASR benefits more through FLOPs reduction and composability with decoder-side compression. These results suggest that Whisper's standard audio-token interface is over-provisioned for many downstream uses.
\section*{Limitations}
stride-$k$ subsampling assumes an encoder that processes speech as overlapping windowed frames, as in Whisper's log-mel frontend, where adjacent tokens share acoustic support by construction. Raw-waveform encoders such as HuBERT and wav2vec~2.0 form their representations differently, and whether a comparable redundancy can be exploited there is an open question we do not address here. The method also operates inside the encoder, applying stride-$k$ between the conv stem and the encoder transformer. This requires access to the encoder's internal computation, so our evaluation covers open-weight ASR models and SpeechLMs; closed models exposed only through an API cannot be modified in this way and fall outside our scope. Finally, we study stride-$k$ purely as a train-free operation and do not retrain or fine-tune any model, a choice made under our compute budget. We view this as a starting point rather than a boundary. Our results show that subsampling is most reliably absorbed by systems that already perform well, which suggests that the redundancy stride-$k$ exploits is partly shaped by training. Whether this redundancy can be deliberately encouraged—so that models are trained to tolerate, or even expect, subsampled token sequences—is a direction we find promising and leave to future work.

\section*{Acknowledgments}
This research was supported by Basic Science Research Program through the National Research Foundation of Korea(NRF) funded by the Ministry of Education (RS-2025-25434151) and the Institute of Information \& Communications Technology Planning \& Evaluation (IITP) grant funded by the Korea government (MSIT) [RS-2021-II211341, Artificial Intelligence Graduate School Program (Chung-Ang University)].

\bibliography{custom}
\appendix

\section*{Usage of Large Language Models}
The manuscript was drafted by the authors. ChatGPT and Gemini-3 were used only for overall language refinement.
\section{About Overlap}
\label{app:overlap}
\subsection{Preliminary for encoder input k}

In speech processing, Short-Time Fourier Transform (STFT) commonly employs Hann windows together with overlapping frame analysis. The overlap is introduced to alleviate abrupt discontinuities between neighboring analysis frames, which would otherwise produce spectral artifacts and instability in the frequency domain. Whisper similarly performs STFT using a 25\,ms Hann window with a 10\,ms hop size. Consequently, adjacent acoustic frames already contain substantial redundant information due to the overlap introduced during feature extraction.

Let the pre-convolutional STFT representations be denoted as $\{f_1,\dots,f_{3000}\}$, where each frame corresponds to a 25\,ms acoustic segment sampled every 10\,ms. Since adjacent STFT frames overlap by 15\,ms, neighboring frames share approximately
\[
\frac{25-10}{25}=0.6
\]
or $60\%$ of their temporal support.

The Whisper convolutional stem consists of two convolutional layers with kernel sizes of $3$, where the second layer applies stride-$2$ subsampling. Consequently, each post-convolutional encoder representation $\{F_1,\dots,F_{1500}\}$ aggregates information from approximately five neighboring STFT frames. For example,
\[
F_1 \rightarrow \{f_1,\dots,f_5\},
\]
\[
F_2 \rightarrow \{f_3,\dots,f_7\},
\]
\[
F_3 \rightarrow \{f_5,\dots,f_9\}.
\]

Considering the 25\,ms STFT window and 10\,ms frame hop, each encoder token therefore spans an effective temporal receptive field of
\[
R = W + (r-1)H,
\]
where $W$ denotes the STFT window size, $H$ the hop size, and $r$ the number of aggregated STFT frames. Substituting Whisper's configuration:
\[
R = 25 + (5-1)\times10 = 65\,\mathrm{ms}.
\]

Furthermore, since the convolutional stem outputs encoder tokens every $20$\,ms due to stride-$2$ downsampling, neighboring encoder representations are already densely overlapped along the temporal axis. Based on this observation, we hypothesize that moderate temporal subsampling can reduce computational redundancy while still preserving near-continuous acoustic coverage. Therefore, we perform deterministic stride-$k$ token pruning before the encoder Transformer computation while explicitly analyzing how much temporal overlap or coverage loss is introduced for different values of $k$.

\subsection{Overlap \& Loss Calculation}

After the convolutional stem, neighboring encoder tokens are spaced by
\[
S = 20\,\mathrm{ms}.
\]

Applying stride-$k$ subsampling retains one token every $k$ encoder tokens, resulting in a retained-token spacing of
\[
\Delta_k = kS = 20k\,\mathrm{ms}.
\]

Since each encoder token spans an effective temporal receptive field of $R=65$\,ms, the temporal overlap between neighboring retained representations can be expressed as
\[
L_{\mathrm{overlap}}(k)
=
\max(0, R-\Delta_k).
\]

Substituting the Whisper configuration yields
\[
L_{\mathrm{overlap}}(k)
=
\max(0, 65-20k).
\]

Let $N_k$ denote the number of retained encoder tokens after stride-$k$ subsampling:
\[
N_k \approx \frac{1500}{k}.
\]

Since neighboring retained encoder representations overlap repeatedly throughout the entire 30-second Whisper input, we first compute the accumulated temporal overlap as
\[
A_{\mathrm{overlap}}(k)
=
L_{\mathrm{overlap}}(k)\cdot(N_k-1).
\]

However, because this accumulated quantity can exceed the original input duration when the encoder sequence is highly redundant, we report a capped global overlap ratio whose maximum value is $100\%$:
\[
O_{\mathrm{global}}(k)
=
\min\left(
1,
\frac{A_{\mathrm{overlap}}(k)}{30000}
\right).
\]

Here, $30000$ corresponds to the 30-second Whisper input duration in milliseconds. This capped ratio should be interpreted as the degree to which temporal redundancy remains over the full input sequence, normalized so that complete temporal overlap coverage is bounded by $100\%$.

If the retained-token spacing exceeds the receptive field size ($\Delta_k > R$), uncovered temporal regions emerge between neighboring retained representations. The temporal gap between neighboring retained representations can then be expressed as
\[
L_{\mathrm{gap}}(k)
=
\max(0,\Delta_k-R).
\]

Substituting the Whisper configuration:
\[
L_{\mathrm{gap}}(k)
=
\max(0,20k-65).
\]

Similarly, we define the accumulated temporal coverage loss as
\[
A_{\mathrm{loss}}(k)
=
L_{\mathrm{gap}}(k)\cdot(N_k-1).
\]

Finally, the normalized global coverage loss ratio becomes
\[
\mathrm{Loss}(k)
=
\frac{
A_{\mathrm{loss}}(k)
}{30000}.
\]

These equations allow us to estimate how much temporal redundancy remains across the entire encoder sequence after subsampling and at which stride factors explicit temporal coverage gaps begin to emerge.

\subsection{Result for Different Values of $k$}

Table~\ref{tab:overlap_loss} summarizes the theoretical capped global temporal overlap and accumulated coverage loss under different stride-$k$ subsampling factors.

\begin{table}[h]
\centering
\small

\begin{tabular}{c c c c}
\toprule
$k$ & Retained Tokens & Overlap Ratio & Loss Ratio \\
\midrule
1 & 1500 & 100.0\% & 0.0\% \\
2 & 750  & 62.4\%  & 0.0\% \\
3 & 500  & 8.3\%   & 0.0\% \\
4 & 375  & 0.0\%   & 18.7\% \\
5 & 300  & 0.0\%   & 34.9\% \\
6 & 250  & 0.0\%   & 45.7\% \\
7 & 215  & 0.0\%   & 53.5\% \\
8 & 188  & 0.0\%   & 59.2\% \\
\bottomrule
\end{tabular}

\caption{Theoretical capped global temporal overlap and accumulated coverage loss under stride-$k$ subsampling. The overlap ratio is capped at $100\%$ to avoid values exceeding the original 30-second input duration.}
\label{tab:overlap_loss}

\end{table}

As shown in Table~\ref{tab:overlap_loss}, the original Whisper encoder representations ($k=1$) already provide complete temporal overlap coverage under the capped global overlap measure, reflecting the highly redundant temporal sampling induced by overlapping STFT analysis and convolutional receptive-field aggregation. Even after removing half of the encoder tokens ($k=2$), substantial temporal redundancy still remains, with approximately $62.4\%$ capped global temporal overlap preserved across the full 30-second encoder sequence.

In contrast, the capped global overlap rapidly decreases at $k=3$ ($8.3\%$), suggesting that stride-$3$ subsampling begins transitioning from redundancy reduction toward actual information removal. Furthermore, explicit temporal coverage loss first emerges at $k\geq4$, where the spacing between retained encoder representations exceeds their effective receptive field.

These observations suggest that Whisper encoder representations are substantially oversampled along the temporal axis. Consequently, moderate temporal subsampling factors, particularly $k=2$, can effectively reduce computational redundancy while still preserving dense temporal acoustic coverage across the encoder sequence.
\section{stride-$k$ Subsampling Algorithm}
\label{app:stridekdetail}

\begin{algorithm}[]
\caption{Stride-$k$ subsampling (\textit{input-side})}
\label{alg:stridek-input}
\begin{algorithmic}[1]
\STATE $H \leftarrow \text{ConvStem}(X)$
\STATE $H \leftarrow H[{::}k,\,:]$
\STATE $H_\text{enc} \leftarrow \text{Encoder}(H)$
\end{algorithmic}
\end{algorithm}

\begin{algorithm}[]
\caption{Stride-$k$ subsampling (output-side)}
\label{alg:stridek-output}
\begin{algorithmic}[1]
\STATE $H \leftarrow \text{ConvStem}(X)$
\STATE $H_\text{enc} \leftarrow \text{Encoder}(H)$
\STATE $H_\text{enc} \leftarrow H_\text{enc}[{::}k,\,:]$ 
\end{algorithmic}
\end{algorithm}

\section{Details set up for Section 4}
Unless otherwise specified, all experiments in this appendix(Section4, CKA, Compound set, SpeechPrune) were conducted with a fixed random seed of 42. We used Python 3.10, PyTorch 2.3.1, torchaudio 2.3.1, and NumPy 1.26.4. Experiments were run on an NVIDIA RTX 6000 Ada GPU. The same random seed was used across sampling and model evaluation procedures to ensure reproducibility.
\label{app:diagonisesetup}
\paragraph{Datasets and model checkpoints.}\mbox{}\\
For the diagnosing experiment, we evaluated stride-$k$ subsampling on three ASR datasets: LibriTTS \texttt{test-clean}, the English subset of ESD, and Common Voice. LibriTTS was used as a clean read-speech benchmark, ESD as an emotionally expressive speech benchmark, and Common Voice as a crowd-sourced benchmark with more diverse recording conditions. For LibriTTS, each audio file was paired with its corresponding \texttt{.normalized.txt} transcript. For ESD, each waveform was paired with its corresponding \texttt{.lab} transcript. For Common Voice, we used the \texttt{test.tsv} metadata file and loaded audio files from the corresponding \texttt{clips} directory. All models were evaluated in frozen mode without training, fine-tuning, or any auxiliary selection module.
\paragraph{Subsampling positions.}\mbox{}\\
We evaluated stride-$k$ subsampling at two positions in the Whisper encoder. For \textit{input-side} subsampling, the log-mel input was first passed through the Whisper convolutional stem, consisting of \texttt{conv1 + GELU} and \texttt{conv2 + GELU}. The resulting hidden sequence was then subsampled as $H’$=\texttt{H[::k,:]} before adding positional embeddings and passing it through the encoder transformer. Thus, the encoder transformer processed a shorter sequence of length approximately $1500/k$. For output-side subsampling, the full 1500-token sequence was first processed by the encoder transformer, and the final encoder hidden states were subsampled as $H’$=\texttt{H[::k,:]} before being passed to the decoder cross-attention. Therefore, input-side subsampling reduces the sequence length processed by both the encoder transformer and decoder cross-attention, whereas output-side subsampling leaves encoder computation unchanged and reduces only the encoder-output sequence consumed by the decoder.

\paragraph{Inference and decoding.}\mbox{}\\
For each audio sample, we constructed Whisper input features using the corresponding HuggingFace \texttt{AutoProcessor}. Audio was loaded with \texttt{torchaudio}, and resampled to the Whisper feature extractor sampling rate. Hypotheses and references were normalized before metric computation by lowercasing and removing non-alphanumeric punctuation for English.

\paragraph{WER and runtime measurement.}\mbox{}\\
ASR performance was measured using Word Error Rate (WER) with \texttt{jiwer}. Runtime was separated into encoder time and decoder time. CUDA synchronization was applied before and after timed regions to obtain stable latency measurements. 

\paragraph{FLOPs estimation.}\mbox{}\\
We estimate the reported total FLOPs as the sum of encoder FLOPs and decoder cross-attention FLOPs:
\[
\mathrm{FLOPs}_{\mathrm{total}}
=
\mathrm{FLOPs}_{\mathrm{enc}}
+
\mathrm{FLOPs}_{\mathrm{cross}}.
\]
Let $T$ be the sequence length processed by the encoder transformer, 
$T_{\mathrm{enc}}$ the encoder-output length passed to decoder cross-attention,
$T_{\mathrm{dec}}$ the generated decoder length, $d$ the hidden size, and
$L_{\mathrm{enc}}$ and $L_{\mathrm{dec}}$ the number of encoder and decoder layers.
Encoder FLOPs are estimated as
\[
\mathrm{FLOPs}_{\mathrm{enc}}
=
2L_{\mathrm{enc}}
\left(
12Td^2 + 2T^2d
\right).
\]
Decoder cross-attention FLOPs are estimated as
\[
\resizebox{0.95\columnwidth}{!}{$
\displaystyle
\mathrm{FLOPs}_{\mathrm{cross}}
=
2L_{\mathrm{dec}}
\left(
(2T_{\mathrm{enc}} + 2T_{\mathrm{dec}})d^2
+
2T_{\mathrm{dec}}T_{\mathrm{enc}}d
\right)
$}
\]
For the baseline, $T=T_{\mathrm{enc}}=1500$. For input-side subsampling with stride $k$, 
$T=T_{\mathrm{enc}}=\lceil1500/k\rceil$. For output-side subsampling, 
$T=1500$ while $T_{\mathrm{enc}}=\lceil1500/k\rceil$. We use the actual generated sequence length as $T_{\mathrm{dec}}$.

\section{Details Result for Section 4}
\label{app:detailResultDiagonise}

Refer to Table~\ref{tab:wer_input_full_2} and \ref{tab:wer_output_full_3}.

\section{Details set up for CKA}
\label{app:ckasetupDetail}

\paragraph{Adjacent-token CKA setup.}\mbox{}\\
For the adjacent-token CKA analysis, we measured representational similarity at two positions in the Whisper encoder: the conv stem output and the final encoder output. The conv stem output was extracted after \texttt{conv1 + GELU} and \texttt{conv2 + GELU}, before positional embedding and transformer layers, while the final encoder output was extracted after the full encoder transformer. We used LibriTTS \texttt{test-clean} and the HuggingFace Whisper checkpoints \texttt{openai/whisper-\{tiny,base,small,medium,\mbox{}\\large-v3\}}. From LibriTTS \texttt{test-clean}, we randomly selected 10 random anchor positions per utterance from the Whisper encoder time axis. Since Whisper produces 1,500 encoder frames, anchor positions were sampled only from valid indices satisfying $t + D_{\max} < 1500$. We used the same randomly sampled anchor positions for all Whisper checkpoints and for both measurement positions, ensuring that the conv stem and encoder output results are directly comparable.

\paragraph{CKA computation.}\mbox{}\\
For each position and each distance $d$, we collected anchor representations $H_t$ and offset representations $H_{t+d}$ and stacked them into two matrices, $X$ and $Y_d$. We then computed linear centered kernel alignment (CKA) as
\[
\mathrm{CKA}(X,Y_d)
=
\frac{\|\tilde{X}^{\top}\tilde{Y}_d\|_F^2}
{\|\tilde{X}^{\top}\tilde{X}\|_F
 \|\tilde{Y}_d^{\top}\tilde{Y}_d\|_F},
\]
where $\tilde{X}$ and $\tilde{Y}_d$ denote sample-centered representations. CKA was computed independently for each Whisper checkpoint, each measurement position, and each token distance. In the final encoder output experiment, we evaluated distances $d \in \{1,2,3,4,5\}$; in the conv stem experiment, we used the same manifest-based sampling procedure and evaluated distance-wise CKA before the transformer layers. All models were evaluated in frozen mode with gradient computation disabled, using \texttt{bfloat16} on CUDA when available and \texttt{float32} otherwise.

\section{Details Result for CKA}
\label{app:cka_full}
Table~\ref{tab:cka_drop} reports the full linear CKA values and
cumulative drops for the conv stem and encoder outputs across all
five Whisper scales at frame distances $d = 1$--$5$, measured on
LibriTTS.

\section{Details set up for Compound set}
\label{app:Compundsetup}
\paragraph{SpeechLM benchmarks and models.}\mbox{}\\
For the SpeechLM compound stride-2 experiment, we evaluated three Whisper-based SpeechLMs: Audio Flamingo 3-7b, Qwen2-Audio-7b-instruct, and LLaMA-Omni2-7b. The experiments were conducted on MMSU and MMAU. For MMSU, we used the HuggingFace dataset \texttt{ddwang2000/MMSU} with the default split used in the evaluation scripts. For MMAU, we used \texttt{AudioLLMs/MMAU}. Each model was evaluated in frozen mode without any training, fine-tuning, or calibration after subsampling. The maximum number of generated tokens was set to 64 and used greedy decoding with \texttt{do\_sample=False}. The reported score is task accuracy, computed by comparing the parsed model answer with the ground-truth option.

\paragraph{Compound stride-2 configuration.}\mbox{}\\
For each model and benchmark, we compared the unmodified baseline against compound stride-2. The baseline processes the full 1,500-token speech encoder sequence. In the compound setting, stride-2 subsampling is applied at two positions: first immediately after the convolutional stem and before the speech encoder transformer, and second immediately after the speech encoder transformer and before the projector or pooling module. This corresponds to $k_{\mathrm{enc\_in}}=2$ and $k_{\mathrm{enc\_out}}=2$. For a 30-second padded audio input, the first subsampling reduces the transformer input from 1,500 to 750 tokens, and the second subsampling reduces the encoder output from 750 to 375 tokens before downstream pooling or projection.

\paragraph{Model-specific audio-token pipelines.}\mbox{}\\
The three SpeechLMs share the same high-level subsampling strategy but differ in how encoder outputs are converted into LLM input tokens. In Audio Flamingo 3 and Qwen2-Audio, the audio pipeline maps the padded mel input to 1,500 conv-stem tokens, processes them with a 32-layer speech encoder transformer, and then applies \texttt{AvgPool1d(2, stride=2)} followed by a projection or normalization module. Thus, the baseline produces 750 audio tokens for the LLM, while compound stride-2 produces 187 tokens after the both subsampling and average pooling, depending on the exact integer rounding. In LLaMA-Omni2, the speech encoder follows the Whisper large-v3-style encoder, but the projector does not use average pooling. Instead, it uses an \texttt{EncoderProjectorConcat} module with a frame-concatenation factor of 5. Therefore, the baseline maps 1,500 speech encoder tokens to 300 LLM-side audio tokens, while compound stride-2 maps 375 tokens to 75 LLM-side audio tokens.

\paragraph{Latency and FLOPs measurement.}\mbox{}\\
For each sample, we measured speech-encoder time and generation time separately, and reported end-to-end latency as their sum. FLOPs were computed analytically from the actual model architecture read from each loaded checkpoint. The speech-side FLOPs include the convolutional stem, the speech encoder transformer, and the audio projector. The convolutional stem always processes the full padded mel input and is therefore unchanged by subsampling. The transformer FLOPs scale with the sequence length after input-side subsampling, and the projector FLOPs scale with the number of tokens after output-side subsampling and pooling or projection. The implementation also estimates LLM prefill and decode FLOPs from the LLM hidden size, number of layers, number of attention heads, number of key-value heads, and intermediate dimension. Total GFLOPs are computed as the sum of speech encoder/projector FLOPs and LLM prefill/decode FLOPs.

\paragraph{Detailed Results.}
\mbox{}\\Table~\ref{tab:compound_full} reports the full compound stride-2
results ($k_{\text{enc\_in}} = k_{\text{enc\_out}} = 2$) for all
datasets and Whisper scales, with per-cell deltas from the
$k = 1$ baseline.

\section{ASR Noise Robustness}
\label{app:noise}

\paragraph{Setup.}
Section~\ref{sec:final} infers the effect of input difficulty by comparing corpora, so difficulty is confounded with domain, speaker population, and recording condition. To isolate difficulty itself, we hold the corpus fixed and degrade it in a controlled way: we add white Gaussian noise to clean LibriTTS test-clean at matched SNR levels. For an utterance with signal power $p_s$, the noise is drawn from $\mathcal{N}(0, p_n)$ with
\[
p_n = p_s / 10^{\mathrm{SNR}/10},
\]
evaluated at $\mathrm{SNR} \in \{\infty, 20, 10, 5, 0\}$\,dB, where $\infty$ denotes the unmodified clean condition. We use 200 utterances sampled with a fixed seed, and the same utterances and the same noise realizations are used for every model scale and every stride condition, so differences across cells reflect the operator rather than sampling. Each SNR level is evaluated under four conditions: the unmodified baseline ($k{=}1$), input-side $k=2$, output-side $k=2$, and compound stride-2. Every condition is compared against the baseline \textit{at the same SNR}, so the reported gap measures the cost of subsampling and not the cost of noise.

\paragraph{Results.}
Table~\ref{tab:noise} reports WER for Whisper-Small and Whisper-Large-v3. Three observations follow.

First, input-side and compound stride-2 degrade monotonically as SNR falls, and more steeply for the smaller model. Compound stride-2 costs $+3.7$ WER on clean speech and $+57.5$ at 0\,dB for Small, against $+0.9$ and $+28.6$ for Large-v3. The capacity ordering of Section~\ref{sec:final} therefore reproduces along an axis where input difficulty is manipulated directly, which supports the interpretation that the ordering reflects a robustness margin rather than a property of any particular corpus.

Second, output-side stride-2 stays within about one point of baseline at every SNR on both scales, and is occasionally slightly better than baseline ($-0.11$ on clean Small, $-0.09$ at 10\,dB). The input/output asymmetry established in Section~\ref{sec4} is thus not an artifact of clean read speech: the redundancy exposed at the encoder output survives acoustic degradation, whereas the receptive-field overlap exploited at the input side does not.

Third, the baseline itself degrades substantially under noise, from $4.72$ to $27.73$ on Small and from $4.21$ to $15.34$ on Large-v3. The widening gap is therefore not a fixed additive cost that noise adds to subsampling, but the consequence of a shrinking redundancy margin: as the signal carries less recoverable information, the fraction of it that stride-$k$ can discard without loss falls.

\paragraph{Error analysis.}
Inspecting utterances that the baseline transcribes correctly but compound stride-2 does not, the dominant added errors are omissions and substitutions of function words, chiefly articles and plural suffixes, rather than substitutions of content words. Two properties plausibly explain this. These units are phonetically short and low in energy, so they are more likely to fall between retained tokens once the temporal sampling rate is halved; and they are recoverable mainly from sentence-level context, so even partial acoustic loss is disproportionately damaging when the decoder cannot fall back on a strong acoustic cue. Under noise these units are already degraded, and stride-2 compounds the loss. Consistently with this account, the same linguistic categories degrade less in the SpeechLM setting, where a downstream language model can restore such words from context, which mirrors the observation in Appendix~\ref{app:failure} that stride-2 is most damaging where the answer depends on evidence the model cannot reconstruct.

\begin{table*}[t]
\centering
\small
\resizebox{\textwidth}{!}{%
\begin{tabular}{lrrrrrrrr}
\toprule
& \multicolumn{4}{c}{\textbf{Whisper-Small}} & \multicolumn{4}{c}{\textbf{Whisper-Large-v3}} \\
\cmidrule(lr){2-5}\cmidrule(lr){6-9}
\textbf{SNR} & \textbf{Base.} & \textbf{In 2} & \textbf{Out 2} & \textbf{Both 2} & \textbf{Base.} & \textbf{In 2} & \textbf{Out 2} & \textbf{Both 2} \\
\midrule
clean & 4.72 & 7.44 \small($+2.72$) & 4.61 \small($-0.11$) & 8.38 \small($+3.66$) & 4.21 & 4.75 \small($+0.54$) & 4.23 \small($+0.03$) & 5.12 \small($+0.92$) \\
20\,dB & 4.84 & 16.80 \small($+11.96$) & 4.89 \small($+0.06$) & 17.37 \small($+12.53$) & 4.38 & 5.64 \small($+1.26$) & 4.46 \small($+0.09$) & 6.81 \small($+2.43$) \\
10\,dB & 7.87 & 19.89 \small($+12.02$) & 7.78 \small($-0.09$) & 28.21 \small($+20.34$) & 5.04 & 10.16 \small($+5.12$) & 5.26 \small($+0.23$) & 11.87 \small($+6.84$) \\
5\,dB & 12.65 & 48.30 \small($+35.65$) & 12.99 \small($+0.34$) & 44.26 \small($+31.62$) & 7.32 & 18.51 \small($+11.19$) & 8.04 \small($+0.72$) & 20.49 \small($+13.16$) \\
0\,dB & 27.73 & 88.38 \small($+60.66$) & 28.61 \small($+0.89$) & 85.24 \small($+57.51$) & 15.34 & 39.91 \small($+24.58$) & 16.19 \small($+0.86$) & 43.98 \small($+28.64$) \\
\bottomrule
\end{tabular}
}
\caption{WER (\%) under additive white Gaussian noise on LibriTTS test-clean ($n=200$ utterances per level, identical across scales and conditions). Each cell reports WER with its gap from the baseline \textit{at the same SNR} in parentheses. Output-side stride-2 stays within about one point of baseline at every level, while input-side and compound stride-2 degrade as conditions worsen, more steeply for the smaller model.}
\label{tab:noise}
\end{table*}

\section{Multilingual Evaluation}
\label{app:multilingual}

The evaluation in the main text focuses on English. To examine whether the behavior of stride-$k$ is language-specific, we repeat the evaluation on the Chinese (zh-CN) portion of Common Voice with the same three Whisper scales used in Section~\ref{sec:final}, applying stride-$k$ at the same two positions and using the same test-split protocol. Since Chinese is not word-segmented, word-level tokenization is not meaningful; following common practice for Chinese ASR evaluation, we report Character Error Rate (CER). All numbers in Table~\ref{tab:cv_zh} are CER (\%), and each stride condition is compared against the baseline of the same scale.

The trends reported in the main text hold across languages. Output-side $k=2$ remains stable in Chinese, staying within one point of baseline at every scale and on Large-v3 even improving slightly ($-0.20$), whereas input-side $k=2$ incurs substantially larger CER increases ($+6.01$, $+6.73$, and $+3.78$) and compound stride-2 the largest. This is the same input/output asymmetry observed in Figure~\ref{fig:eachexperimentresult} and discussed in Section~\ref{sec4}, and its reappearance under a different language and a different error metric indicates that the asymmetry follows from where the redundancy is created in Whisper's encoder rather than from properties of English.

The stability of the output side also strengthens with scale, with the gap falling from $+0.67$ on Small to $+0.45$ on Medium and $-0.20$ on Large-v3. The input side is less regular: Large-v3 absorbs input-side $k=2$ best ($+3.78$), but the compound gap stays roughly flat across scales ($+8.30$ to $+9.04$) rather than shrinking as it does on the English benchmarks, so the capacity trend of Section~\ref{sec:final} is weaker here. Absolute CER is also higher than the English WERs in Table~\ref{tab:compound-main}, reflecting the difficulty of the zh-CN portion of Common Voice rather than an effect of subsampling. Broader multilingual evaluation, including Japanese and other Common Voice languages, and an analysis of how the degradation relates to a language's resource level and character-level information density, remain future work.

\begin{table}[t]
\centering
\small
\setlength{\tabcolsep}{4pt}
\begin{tabular}{lrrrr}
\toprule
\textbf{Scale} & \textbf{Base.} & \textbf{In 2} & \textbf{Out 2} & \textbf{Both 2} \\
\midrule
Small & 34.92 & 40.93 & 35.59 & 43.22 \\
 & & \small($+6.01$) & \small($+0.67$) & \small($+8.30$) \\
Medium & 27.22 & 33.95 & 27.67 & 36.35 \\
 & & \small($+6.73$) & \small($+0.45$) & \small($+9.13$) \\
Large-v3 & 16.68 & 20.46 & 16.48 & 25.71 \\
 & & \small($+3.78$) & \small($-0.20$) & \small($+9.04$) \\
\bottomrule
\end{tabular}
\caption{Chinese (Common Voice zh-CN) CER (\%) under stride-2, with the change from each scale's own baseline in parentheses. In~2 and Out~2 denote input-side and output-side stride-2, Both~2 the compound setting. Output-side stride-2 stays within one point of baseline at every scale and slightly improves on Large-v3.}
\label{tab:cv_zh}
\end{table}

\section{Generalization Beyond Whisper}
\label{app:beyondwhisper}

Section~\ref{sec:final} and the Limitations discuss stride-$k$ as an operation on Whisper. This appendix makes the boundary of that claim precise. We do not claim that stride-$k$ generalizes to all speech encoders: the method was motivated by, and is introduced as exploiting, the specific preprocessing redundancy of Whisper's log-mel frontend and convolutional stem. What we establish here is narrower and more useful, namely that the operative condition is not the Whisper architecture itself but whether the preprocessing pipeline produces sufficient temporal overlap between adjacent tokens, so that encoders sharing this property inherit the operation while encoders lacking it do not. We state the condition (Appendix~\ref{app:beyondwhisper:rf}), test it on encoders that violate it (Appendix~\ref{app:beyondwhisper:ssl}) and on one that satisfies it (Appendix~\ref{app:beyondwhisper:phi4}), and connect the outcome to the similarity analysis of Section~\ref{sec:cka} (Appendix~\ref{app:beyondwhisper:cka}).

Following Appendix~\ref{app:overlap}, a post-frontend token spans a receptive field $R$ and is emitted every $S$ ms, so retaining every $k$-th token leaves an overlap $L_\text{overlap}(k)=\max(0, R-kS)$ and a coverage gap $L_\text{gap}(k)=\max(0, kS-R)$. Stride-$k$ removes redundancy rather than signal exactly when $L_\text{gap}(k)=0$, which for $k=2$ requires $R \ge 2S$. This is a property of the frontend alone: it is fixed by the analysis window, the hop, and the strides of the convolutional stem, and it can therefore be computed for any encoder without running it.

\subsection{Receptive fields of three frontends}
\label{app:beyondwhisper:rf}

We apply the construction $R = W + (r-1)H$ of Appendix~\ref{app:overlap}, where $W$ is the analysis window, $H$ the hop, and $r$ the number of frames a post-stem token aggregates.

\begin{itemize}[leftmargin=*,itemsep=3pt,topsep=3pt]
\item \textbf{Whisper.} Log-mel with $W=25$\,ms and $H=10$\,ms; \texttt{conv1} ($k{=}3$, $s{=}1$) followed by \texttt{conv2} ($k{=}3$, $s{=}2$) aggregates $r=5$ mel frames, giving $R = 25 + 4\times 10 = 65$\,ms at $S=20$\,ms. The overlap originates before the stem: adjacent mel frames already share $(25-10)/25 = 60\%$ of their temporal support, and the stem compounds this by aggregating five such frames into every token.
\item \textbf{Phi-4-multimodal conformer.} Log-mel with $W=400$ samples ($25$\,ms) and $H=160$ samples ($10$\,ms); the \texttt{NemoConvSubsampling} stem (\texttt{dw\_striding}, time reduction $8$, i.e.\ three depthwise-separable \texttt{Conv2d} layers of kernel $3$ and stride $2$) spans $r=15$ mel frames, giving $R = 400 + 14\times 160 = 2640$ samples $=165$\,ms at $S = 8\times 160 = 1280$ samples $=80$\,ms. The pipeline is Whisper-like, mel extraction followed by a convolutional stem, so adjacent post-stem tokens overlap by construction, but the margin is thinner in relative terms.
\item \textbf{Raw-waveform encoders.} wav2vec~2.0, HuBERT, WavLM, and wav2vec2-conformer share a seven-layer convolutional feature extractor (kernels $[10,3,3,3,3,2,2]$, strides $[5,2,2,2,2,2,2]$) applied directly to the waveform, skipping the mel transformation entirely. This gives $R = 400$ samples $=25$\,ms at $S = 320$ samples $=20$\,ms, verified empirically in that a $16{,}000$-sample input yields $49$ frames. The receptive field is barely wider than the spacing, so stride-2 places retained tokens $40$\,ms apart while each covers only $25$\,ms, leaving a $15$\,ms gap in every $40$\,ms of timeline, or roughly $37\%$ of the signal outside any retained token's receptive field.
\end{itemize}

Table~\ref{tab:frontend_rf} summarizes the resulting stride-2 behavior. The two mel-plus-convolution frontends retain positive overlap and open no gap, and their ordering is itself informative: the Phi-4-multimodal conformer inherits the property but with a thinner margin ($L_\text{overlap}=5$\,ms against Whisper's $25$\,ms), which predicts that it should tolerate stride-2 while degrading more than Whisper does. Raw-waveform frontends open a gap immediately, so for them stride-2 is not redundancy removal but signal removal.

\begin{table}[t]
\centering
\small
\setlength{\tabcolsep}{3pt}
\resizebox{\columnwidth}{!}{%
\begin{tabular}{lrrrrr}
\toprule
\textbf{Frontend} & $R$ & $S$ & \textbf{Overlap} & $L_\text{ovl}$ & $L_\text{gap}$ \\
\midrule
Whisper (mel + conv stem) & 65 & 20 & 45 (69\%) & 25 & 0 \\
Phi-4-MM (mel + conv sub.) & 165 & 80 & 85 (52\%) & 5 & 0 \\
Raw-waveform conv & 25 & 20 & 5 (20\%) & 0 & 15 \\
\bottomrule
\end{tabular}
}
\caption{Token receptive field $R$, spacing $S$, and adjacent overlap $R-S$ before the encoder transformer, with the residual overlap $L_\text{ovl}$ and coverage gap $L_\text{gap}$ under stride-2 (all in ms). Raw-waveform conv covers wav2vec~2.0, HuBERT, WavLM, and wav2vec2-conformer, which share one convolutional feature extractor.}
\label{tab:frontend_rf}
\end{table}

\subsection{Stride-\texorpdfstring{$k$}{k} on raw-waveform encoders}
\label{app:beyondwhisper:ssl}

We apply input-side, output-side, and compound stride to four raw-waveform encoders decoded with CTC on LibriTTS test-clean, using the same corpus, normalization, and WER computation as Section~\ref{sec4}. Table~\ref{tab:ssl_ctc} reports the result: every model degrades sharply at $k=2$ at either position, and compound stride-2 destroys recognition entirely. There is no regime analogous to Whisper's, where $k=2$ is absorbed at both positions; the collapse is immediate, as the coverage analysis predicts.

\begin{table}[t]
\centering
\small
\setlength{\tabcolsep}{4pt}
\begin{tabular}{lrrrrr}
\toprule
\textbf{Model} & \textbf{Base.} & \textbf{In 2} & \textbf{In 3} & \textbf{Out 2} & \textbf{Both 2} \\
\midrule
wav2vec~2.0 & 3.06 & 17.21 & 73.58 & 89.53 & 97.77 \\
HuBERT & 3.68 & 24.55 & 76.32 & 81.05 & 97.50 \\
WavLM & 7.60 & 48.67 & 92.00 & 81.03 & 97.79 \\
\bottomrule
\end{tabular}
\caption{WER (\%) under stride-$k$ for raw-waveform encoders with CTC decoding on LibriTTS test-clean ($n=4837$). In~$k$ and Out~$k$ denote input-side and output-side stride, Both~2 the compound setting.}
\label{tab:ssl_ctc}
\end{table}

\subsection{Stride-2 on a mel-input conformer}
\label{app:beyondwhisper:phi4}

Phi-4-multimodal pairs conformer encoder blocks with the mel-plus-convolution frontend of Appendix~\ref{app:beyondwhisper:rf}, so the condition predicts that it should tolerate compound stride-2. We evaluate it on MMSU ($n=5000$) and MMAU ($n=1000$) under the protocol of Appendix~\ref{app:Compundsetup}. It degrades rather than collapsing: MMAU accuracy moves from $62.20$ to $56.50$ ($-5.70$), MMSU Perception from $37.13$ to $32.52$ ($-4.61$), and MMSU Reasoning from $70.41$ to $58.97$ ($-11.45$), while the audio-token count falls from $170.9$ to $43.3$ on MMAU.

Table~\ref{tab:phi4_delta} places these deltas beside the three Whisper-based SpeechLMs of Table~\ref{tab:compound-main}. Two things follow. First, the qualitative behavior matches the prediction: an encoder that satisfies the coverage condition remains usable under a $75\%$ token reduction, in sharp contrast to the raw-waveform encoders of Table~\ref{tab:ssl_ctc}. Second, the drop is larger than for any Whisper-based model, which is consistent with the thinner overlap margin computed in Table~\ref{tab:frontend_rf} ($52\%$ against Whisper's $69\%$). The condition therefore predicts not only whether stride-2 transfers but, in the right direction, how much it costs.

\begin{table}[t]
\centering
\small
\resizebox{\columnwidth}{!}{%
\begin{tabular}{lrr}
\toprule
\textbf{Model} & \textbf{Perception} $\Delta$ & \textbf{Reasoning} $\Delta$ \\
\midrule
Audio Flamingo 3 & $-1.00$ & $-2.81$ \\
Qwen2-Audio & $-1.62$ & $-6.16$ \\
LLaMA-Omni~2 & $-3.06$ & $-9.71$ \\
\midrule
Phi-4-MM (mel conformer) & $-4.61$ & $-11.45$ \\
\bottomrule
\end{tabular}
}
\caption{MMSU accuracy change under compound stride-2. The first three rows are the Whisper-based SpeechLMs of Table~\ref{tab:compound-main}; the last is the mel-input conformer of Phi-4-multimodal, which degrades more than any Whisper-based model but does not collapse.}
\label{tab:phi4_delta}
\end{table}

\subsection{Adjacent-token similarity for raw-waveform encoders}
\label{app:beyondwhisper:cka}

The receptive-field argument is geometric and holds before any weights are consulted. Section~\ref{sec:cka} showed that the same redundancy is visible representationally, as high similarity between adjacent tokens. If the frontend account is correct, raw-waveform encoders should show that signature much more weakly. We therefore repeat the measurement of Section~\ref{sec:cka} on the self-supervised encoders. As noted there and in Appendix~\ref{app:ckasetupDetail}, absolute CKA is not comparable \textit{across} positions, so we compare the same position type across encoders rather than the two positions against each other.

Table~\ref{tab:ssl_cka} reports the result. At the convolutional output, where the geometric overlap resides, Whisper reaches $0.964$ at $d=1$ while the self-supervised encoders reach only $0.582$--$0.755$, and their similarity falls off far faster with distance ($0.207$--$0.287$ at $d=3$ against Whisper's $0.851$). The representational deficit tracks the receptive-field deficit, and predicts the collapse in Table~\ref{tab:ssl_ctc} from an independent measurement: the tokens stride-$k$ retains are poor substitutes for the ones it drops.

Taken together, these results locate the operative condition. It is not the Whisper architecture, nor the encoder block family, but whether the preprocessing pipeline produces sufficient temporal overlap between adjacent tokens. Overlapping windowed analysis is standard practice in signal processing, adopted precisely to avoid spectral discontinuities between neighboring frames (Appendix~\ref{app:overlap}), so the property stride-$k$ exploits is not idiosyncratic to Whisper. Any encoder that follows this convention with mel extraction and a convolutional stem is a candidate for stride-$k$ subsampling, and Table~\ref{tab:frontend_rf} shows that its viability can be assessed from the frontend configuration alone, before any evaluation is run.

\begin{table}[t]
\centering
\small
\setlength{\tabcolsep}{4pt}
\begin{tabular}{llrrr}
\toprule
\textbf{Model} & \textbf{Position} & $d{=}1$ & $d{=}2$ & $d{=}3$ \\
\midrule
Whisper-Tiny & Conv & 0.964 & 0.904 & 0.851 \\
 & Enc Out & 0.881 & 0.776 & 0.738 \\
\midrule
wav2vec~2.0 & Conv & 0.582 & 0.338 & 0.207 \\
 & Enc Out & 0.038 & 0.043 & 0.021 \\
HuBERT & Conv & 0.755 & 0.481 & 0.287 \\
 & Enc Out & 0.702 & 0.482 & 0.329 \\
WavLM & Conv & 0.743 & 0.469 & 0.280 \\
 & Enc Out & 0.304 & 0.104 & 0.071 \\
\bottomrule
\end{tabular}
\caption{Linear CKA between tokens at frame distance $d$, measured at the convolutional output and the encoder output on LibriTTS. Whisper-Tiny is shown for reference. At the convolutional position the self-supervised encoders show much weaker adjacent similarity than Whisper, mirroring the receptive-field deficit in Table~\ref{tab:frontend_rf}.}
\label{tab:ssl_cka}
\end{table}

\section{Details Result for SpeechPrune}
\label{app:spDetail}

Tables~\ref{tab:mmsu_audioflamingo3}--\ref{tab:mmau_qwenaudio} report
the category-level accuracy, FLOPs, and latency for SpeechPrune (SP),
input-side stride-2 (Enc), and the combined setting (Enc+SP) on MMSU
and MMAU across the three SpeechLMs.

\paragraph{SpeechPrune and combined setting.}\mbox{}\\
We evaluate SpeechPrune as a training-free post-projector speech-token pruning. In the SpeechPrune-only setting, the full audio encoder is executed without subsampling, producing approximately 750 post-projector speech tokens for a 30-second padded input. SpeechPrune is then applied in two phases. First, projected speech tokens are scored by cosine similarity to the question text embeddings, and a frame-wise softmax allocation selects the most relevant tokens within each one-second segment. Second, the remaining tokens are scored using binarized attention computed from the sign-binarized query and key weights of the first LLM layer, and the top-scoring tokens are retained. We evaluate pruning rates of 20\%, 40\%, 60\%, and 80\%.

\paragraph{Ours + SpeechPrune.}\mbox{}\\
For the combined setting, we first apply our input-side stride-2 subsampling before the speech encoder transformer. This reduces the encoder-transformer input from 1,500 to 750 tokens and yields approximately 375 post-projector speech tokens after average pooling and projection. SpeechPrune is then applied to this already-reduced speech-token sequence with the same two-phase procedure. Thus, SpeechPrune-only reduces the number of audio tokens passed to the LLM after the full speech encoder has been computed, whereas the combined method reduces the speech encoder computation first and then further prunes the resulting LLM-side speech tokens. This distinction is used when reporting FLOPs and latency: SpeechPrune-only mainly reduces downstream LLM computation, while the combined setting reduces both speech-encoder computation and downstream LLM input length.

\section{Failure-Case Analysis of Compound Stride-2}
\label{app:failure}
 
This section expands the per-sample failure analysis summarized in Section~\ref{sec:final}. The aggregate results in Table~\ref{tab:compound-main} report how much accuracy changes under compound stride-2; here we instead ask which questions change, and whether the newly induced errors are distributed uniformly across task types or concentrated on particular ones.
 
\subsection{Setup}
\label{app:failure:setup}

For each of the three SpeechLMs (Audio Flamingo~3, Qwen2-Audio, LLaMA-Omni~2), we run inference on the full MMSU and MMAU question sets under two settings: the unmodified baseline ($k\!=\!1$) and compound stride-2 ($k_{\text{enc\_in}}\!=\!k_{\text{enc\_out}}\!=\!2$). A failure case is a question that the baseline answers correctly but compound stride-2 answers incorrectly. These cases are exactly the errors introduced by subsampling, excluding questions that were already wrong at the baseline.

For a task category $c$, we define the failure rate as
\[
\begin{aligned}
\mathrm{FR}(c)
&=
\frac{
\#\{i \in c :
y_i^{\mathrm{base}} = y_i,\;
y_i^{\mathrm{stride}} \ne y_i\}
}{
\#\{i \in c :
y_i^{\mathrm{base}} = y_i\}
}.
\end{aligned}
\]
Here, $y_i$ is the ground-truth answer, and
$y_i^{\mathrm{base}}$ and $y_i^{\mathrm{stride}}$ denote the predictions
under the baseline and compound stride-2 settings, respectively.
Thus, $\mathrm{FR}(c)$ is the fraction of questions a model originally got right in category $c$ that are lost under stride-2. The denominator is the model's own baseline-correct count, so the failure rate is comparable across categories of different size and across models of different baseline accuracy.

Task categories are obtained by grouping questions from their prompt text, following the task taxonomies of MMSU and MMAU. Categories whose failure rates are small and whose question counts are low are merged into an Others row to keep the tables compact; the prosodic category Stress/Emphasis is kept separate despite a moderate rate because it is directly relevant to the temporal-resolution discussion. $N$ denotes the number of baseline-correct questions summed over the three models, i.e., the pooled denominator.
\subsection{Results}
\label{app:failure:results}
 
Tables~\ref{tab:fr-mmsu} and~\ref{tab:fr-mmau} report per-category failure rates for MMSU and MMAU. Three observations follow.
 
\paragraph{The overall failure rate follows the baseline ordering.}
On MMSU the overall failure rate is 9.4\% for AF3, 14.8\% for Qwen2-Audio, and 24.5\% for LLaMA-Omni~2; on MMAU it is 6.8\%, 12.6\%, and 20.6\% respectively. Both benchmarks independently reproduce the ordering AF3 $<$ Qwen2-Audio $<$ LLaMA-Omni~2, matching the accuracy trend in Table~2 and supporting the interpretation that stronger baselines retain a larger margin to absorb the fixed fraction of redundancy removed by subsampling.
 
\paragraph{Temporally fine-grained categories are vulnerable for every baseline.} On MMSU, Intonation (20.4\% averaged over the three models) and Pause (20.3\%) are the two highest non-Others categories, and each exceeds its model's overall failure rate for all three models. On MMAU, Counting (20.7\%) shows the same property, with a strikingly flat profile across baselines (21.4\% / 22.0\% / 18.8\%). These categories share a dependence on temporal structure, the shape of an intonation contour, the position of a pause, the number of events in time, which is precisely what stride-2 compresses by halving the encoder's temporal sampling rate. Their consistency across baselines indicates a property of the operator rather than of any single model.
 
\paragraph{Other categories track baseline strength, not task type.}
Categories such as Pitch/Volume and Transcription on MMSU, or Emotion and SoundRecognition on MMAU, show a steep spread across baselines (e.g.\ MMSU Pitch/Volume: 1.6\% for AF3 vs.\ 38.0\% for LLaMA-Omni~2). Their failure rate is governed by baseline strength rather than by an intrinsic vulnerability of the category, and we therefore do not interpret them as evidence about which acoustic cues stride-2 removes. Only the temporally fine-grained categories above exhibit a baseline-independent pattern.
 
\begin{table}[t]
\centering
\small
\setlength{\tabcolsep}{4pt}
\resizebox{\columnwidth}{!}{%
\begin{tabular}{lcccc c}
\toprule
\multirow{2}{*}{Category} & \multicolumn{3}{c}{Failure rate (\%)} &
\multirow{2}{*}{Avg} & \multirow{2}{*}{$N$} \\
\cmidrule(lr){2-4}
 & AF3 & Qwen2-A. & L-Omni2 & & \\
\midrule
Intonation        & 23.9 & 16.7 & 20.0 & \textbf{20.4} & 181 \\
Pause             & 15.6 & 25.4 & 19.4 & \textbf{20.3} & 316 \\
Syllable          & 16.1 & 12.5 & 32.3 & 19.6 & 102 \\
Poetic            & 19.6 & 28.6 & 10.0 & 19.2 & 104 \\
StopRelease       &  7.9 & 11.3 & 31.0 & 18.1 & 149 \\
Transcription     &  5.9 & 13.6 & 32.9 & 16.3 & 608 \\
Pitch/Volume      &  1.6 &  8.0 & 38.0 & 16.2 & 346 \\
SpeakerTraits     & 11.7 & 18.6 & 19.6 & 16.2 & 2585 \\
Semantics-Reason. &  5.8 & 12.6 & 33.7 & 15.8 & 1777 \\
Stress/Emphasis   & 11.9 & 11.9 & 15.2 & 12.9 & 302 \\
Others            &  8.1 & 10.6 & 17.2 & 11.4 & 1435 \\
\midrule
\textbf{Overall}  & \textbf{9.4} & \textbf{14.8} & \textbf{24.5} & -- & 8205 \\
\bottomrule
\end{tabular}
}
\caption{Per-category failure rate on MMSU: the fraction of
baseline-correct ($k\!=\!1$) questions that become incorrect under
compound stride-2. Rows are sorted by the three-model average; $N$ is
the pooled number of baseline-correct questions. Intonation and Pause
are the highest non-\emph{Others} categories and exceed each model's
overall failure rate.}
\label{tab:fr-mmsu}
\end{table}
 
\begin{table}[t]
\centering
\small
\setlength{\tabcolsep}{4pt}
\resizebox{\columnwidth}{!}{%
\begin{tabular}{lcccc c}
\toprule
\multirow{2}{*}{Category} & \multicolumn{3}{c}{Failure rate (\%)} &
\multirow{2}{*}{Avg} & \multirow{2}{*}{$N$} \\
\cmidrule(lr){2-4}
 & AF3 & Qwen2-A. & L-Omni2 & & \\
\midrule
SpeakerRelation   &  0.0 &  0.0 & 78.6 & 25.0 &  44 \\
Emotion           &  4.8 & 32.0 & 38.5 & 23.7 &  59 \\
SoundRecognition  &  5.7 & 28.2 & 34.9 & 21.5 & 135 \\
Counting          & 21.4 & 22.0 & 18.8 & \textbf{20.7} & 145 \\
Stress/Phoneme    & 10.9 & 25.6 & 31.6 & 20.4 & 108 \\
Others            &  5.3 &  8.0 & 15.9 &  9.2 & 1291 \\
\midrule
\textbf{Overall}  & \textbf{6.8} & \textbf{12.6} & \textbf{20.6} & -- & 1782 \\
\bottomrule
\end{tabular}
}
\caption{Per-category failure rate on MMAU, defined as in
Table~\ref{tab:fr-mmsu}. Counting is the only category with a
baseline-flat profile (21.4\%/22.0\%/18.8\%) and exceeds every model's
overall failure rate; the remaining categories spread widely across
baselines.}
\label{tab:fr-mmau}
\end{table}
 
\subsection{Discussion}
\label{app:failure:discussion}
 
The failure-case analysis refines the aggregate results of Table~2 in two ways. The overall failure rate, reproduced independently on MMSU and MMAU, confirms that the accuracy drops in Table~2 are not driven by a few outlier questions but reflect a baseline-dependent error rate spread across the benchmark. More specifically, the per-category view isolates a baseline-independent effect. Tasks defined by temporal structure, intonation, pause, and counting, are eroded first under stride-2, regardless of which SpeechLM is used. This is the expected signature of an operator that reduces the encoder's temporal sampling rate, and it indicates that the residual cost of compound stride-2 is not uniform but falls disproportionately on temporally fine-grained perception.

\section{Details Result for Distil-Whisper}
\label{app:distilwhisper}
Table~\ref{tab:adistilwhisper} reports the ASR performance and total
FLOPs of Distil-Whisper large-v3 under stride-$k$ subsampling.
We compare the unmodified Distil-Whisper baseline, Distil-Whisper
with input-side stride-2, and Distil-Whisper with compound stride-2.
Applying compound stride-2 reduces total FLOPs by roughly a further
54\% relative to the Distil-Whisper baseline, with WER staying close
to baseline on LibriTTS and ESD and increasing more on the harder
Common Voice.

\input{app_table}

\end{document}

%% file: finalResult.tex
\begin{table*}[t]
\centering
\scriptsize
\setlength{\tabcolsep}{2pt}

\resizebox{\linewidth}{!}{%
\begin{tabular}{ll ccc ccc ccc}
\toprule
\multicolumn{2}{l}{\textbf{Whisper}} & \multicolumn{3}{c}{\textbf{Small}}
  & \multicolumn{3}{c}{\textbf{Medium}}
  & \multicolumn{3}{c}{\textbf{Large}} \\
\cmidrule(lr){3-5} \cmidrule(lr){6-8} \cmidrule(lr){9-11}
 & 
  & \textbf{Base.} & \textbf{Both $k{=}2$} & \textbf{$\Delta$}
  & \textbf{Base.} & \textbf{Both $k{=}2$} & \textbf{$\Delta$}
  & \textbf{Base.} & \textbf{Both $k{=}2$} & \textbf{$\Delta$} \\
\midrule
\multirow{3}{*}{\shortstack[l]{ASR\\(WER \%)}}
  & LibriTTS     & 5.02  & 9.22  & $+$4.20  & 4.61  & 7.01  & $+$2.40 & 4.31  & 5.38  & $+$1.07 \\
  & ESD          & 9.39  & 14.23 & $+$4.84  & 8.65  & 12.16 & $+$3.50 & 9.63  & 10.64 & $+$1.01 \\
  & Common Voice & 17.94 & 31.34 & $+$13.40 & 14.26 & 23.82 & $+$9.56 & 11.66 & 21.56 & $+$9.90 \\
\midrule
\multicolumn{2}{c}{\textbf{GFLOPs}} 
  & 381.51 & 159.45 & $-$58.2\% 
  & 1282.00 & 548.30 & $-$57.2\% 
  & 2577.71 & 1119.27 & $-$56.6\% \\
\multicolumn{2}{c}{\textbf{Total ms}} 
  & 143.76 & 148.88 & $+$3.6\%   
  & 234.85 & 232.25 & $-$1.1\%   
  & 313.94 & 295.17 & $-$6.0\% \\
\bottomrule
\end{tabular}%
}
\vspace{1mm}

\resizebox{\linewidth}{!}{%
\begin{tabular}{ll ccccc ccccc ccccc}
\toprule
\multicolumn{2}{l}{\textbf{SpeechLMs}} 
  & \multicolumn{5}{c}{\textbf{Audio Flamingo3}}
  & \multicolumn{5}{c}{\textbf{Qwen2-Audio}}
  & \multicolumn{5}{c}{\textbf{LLaMA-Omni2}} \\
\cmidrule(lr){3-7} \cmidrule(lr){8-12} \cmidrule(lr){13-17}
 &
  & \textbf{Base.} & \textbf{Both $k{=}2$} & \textbf{$\Delta$} & \textbf{Hurt} & \textbf{Help}
  & \textbf{Base.} & \textbf{Both $k{=}2$} & \textbf{$\Delta$} & \textbf{Hurt} & \textbf{Help}
  & \textbf{Base.} & \textbf{Both $k{=}2$} & \textbf{$\Delta$} & \textbf{Hurt} & \textbf{Help} \\
\midrule

\multirow{5}{*}{\shortstack[l]{MMSU\\Perception}}
  & Ling.Semantics      
  & 52.28 & 51.97 & $-$0.31 & 3.46 & 3.15
  & 53.54 & 51.65 & $-$1.89 & 6.77 & 4.88
  & 31.65 & 28.82 & $-$2.83 & 7.40 & 4.57 \\

  & Ling.Phonology      
  & 37.75 & 37.54 & $-$0.21 & 6.63 & 6.42
  & 37.11 & 36.36 & $-$0.75 & 5.99 & 5.24
  & 33.16 & 28.24 & $-$4.92 & 8.88 & 3.96 \\

  & Para.Speaker Traits 
  & 49.13 & 44.32 & $-$4.81 & 9.17 & 4.37
  & 44.10 & 40.39 & $-$3.71 & 8.73 & 5.02
  & 24.24 & 24.24 & $\phantom{+}$0.00 & 8.08 & 8.08 \\

  & Para.Speaking Style 
  & 34.60 & 34.60 & $\phantom{+}$0.00 & 1.45 & 1.45
  & 31.16 & 30.07 & $-$1.09 & 4.53 & 3.44
  & 27.36 & 24.64 & $-$2.72 & 9.24 & 6.52 \\

\cmidrule(lr){2-17}
  & \textit{Average}    
  & \textit{42.67} & \textit{41.67} & \textit{$-$1.00} & -- & --
  & \textit{41.12} & \textit{39.50} & \textit{$-$1.62} & -- & --
  & \textit{29.96} & \textit{26.90} & \textit{$-$3.06} & -- & -- \\

\midrule

\multirow{5}{*}{\shortstack[l]{MMSU\\Reasoning}}
  & Ling.Semantics      
  & 87.45 & 81.77 & $-$5.68 & 6.86 & 1.17
  & 81.05 & 72.29 & $-$8.76 & 12.09 & 3.34
  & 61.10 & 47.38 & $-$13.72 & 16.52 & 2.80 \\

  & Ling.Phonology      
  & 76.97 & 76.25 & $-$0.72 & 5.02 & 4.30
  & 64.69 & 60.80 & $-$3.89 & 7.98 & 4.09
  & 64.38 & 57.22 & $-$7.16 & 10.44 & 3.28 \\

  & Para.Speaker Traits 
  & 44.25 & 43.36 & $-$0.89 & 6.64 & 5.75
  & 47.79 & 46.02 & $-$1.77 & 8.85 & 7.08
  & 29.20 & 26.99 & $-$2.21 & 8.85 & 6.64 \\

  & Para.Speaking Style 
  & 43.12 & 46.79 & $+$3.67 & 4.59 & 8.26
  & 44.95 & 35.78 & $-$9.17 & 11.01 & 1.83
  & 39.45 & 32.11 & $-$7.34 & 12.84 & 5.50 \\

\cmidrule(lr){2-17}
  & \textit{Average}    
  & \textit{77.19} & \textit{74.38} & \textit{$-$2.81} & -- & --
  & \textit{69.71} & \textit{63.55} & \textit{$-$6.16} & -- & --
  & \textit{58.47} & \textit{48.76} & \textit{$-$9.71} & -- & -- \\

\midrule

\multirow{4}{*}{\shortstack[l]{MMAU\\Audio}}
  & Sound  
  & 73.6 & 72.1 & $-$1.5 & 4.80 & 3.30
  & 59.8 & 55.3 & $-$4.5 & 6.61 & 2.10
  & 52.6 & 45.3 & $-$7.2 & 11.41 & 4.20 \\

  & Music  
  & 79.3 & 78.4 & $-$0.9 & 4.49 & 3.59
  & 55.4 & 53.3 & $-$2.1 & 5.39 & 3.29
  & 52.7 & 53.6 & $+$0.9 & 5.69 & 6.59 \\

  & Speech 
  & 58.6 & 55.9 & $-$2.7 & 5.11 & 2.40
  & 53.8 & 49.5 & $-$4.2 & 9.31 & 5.11
  & 48.9 & 38.7 & $-$10.2 & 14.71 & 4.50 \\

\cmidrule(lr){2-17}
  & \textit{Average}    
  & \textit{70.5} & \textit{68.8} & \textit{$-$1.7} & -- & --
  & \textit{56.3} & \textit{52.7} & \textit{$-$3.6} & -- & --
  & \textit{51.4} & \textit{45.9} & \textit{$-$5.5} & -- & -- \\

\midrule
\multicolumn{2}{c}{\textbf{GFLOPs}}
  & 6163.58 & 2910.34 & $-$52.8\% & -- & --
  & 6226.83 & 2991.48 & $-$51.9\% & -- & --
  & 7449.83 & 3232.28 & $-$56.6\% & -- & -- \\

\multicolumn{2}{c}{\textbf{Total ms}}
  & 122.16 & 94.92 & $-$22.3\% & -- & --
  & 112.25 & 90.27 & $-$19.6\% & -- & --
  & 194.35 & 141.12 & $-$27.4\% & -- & -- \\

\bottomrule
\end{tabular}%
}

\caption{
Compound stride-2 results on Whisper ASR and Whisper-based SpeechLMs.
ASR results are reported in WER, and SpeechLM results in accuracy.
$\Delta$ is computed as compound stride-2 minus baseline.
Hurt and Help indicate baseline-correct/subsampled-incorrect and baseline-incorrect/subsampled-correct cases, respectively.
}
\label{tab:compound-main}
\vspace{-2mm}
\end{table*}

%% file: app_table.tex
\begin{table*}[t]
\centering
\small
\resizebox{\textwidth}{!}{%
\begin{tabular}{lllrrrrrr}
\toprule
\textbf{Dataset} & \textbf{Model} & \textbf{Metric} & $k=1$ & $k=2$ & $k=3$ & $k=4$ & $k=5$ & $k=6$ \\
\midrule
\multirow{10}{*}{LibriTTS} & \multirow{2}{*}{Tiny} & WER (\%) & \textbf{8.86} & 21.23 & 171.1 & 253.8 & 202.8 & 190.3 \\
 & & GFLOPs & \textbf{39\,G (100\%)} & 16\,G (41\%) & 10\,G (26\%) & 7.3\,G (19\%) & 5.7\,G (15\%) & 4.7\,G (12\%) \\
 & \multirow{2}{*}{Base} & WER (\%) & \textbf{6.87} & 12.47 & 73.12 & 190.3 & 183.1 & 154.0 \\
 & & GFLOPs & \textbf{94\,G (100\%)} & 40\,G (43\%) & 25\,G (27\%) & 19\,G (20\%) & 15\,G (16\%) & 12\,G (13\%) \\
 & \multirow{2}{*}{Small} & WER (\%) & \textbf{5.02} & 8.04 & 40.04 & 112.3 & 131.2 & 124.2 \\
 & & GFLOPs & \textbf{382\,G (100\%)} & 171\,G (45\%) & 109\,G (29\%) & 81\,G (21\%) & 64\,G (17\%) & 53\,G (14\%) \\
 & \multirow{2}{*}{Medium} & WER (\%) & \textbf{4.61} & 6.22 & 19.51 & 73.16 & 107.7 & 109.4 \\
 & & GFLOPs & \textbf{1.3k\,G (100\%)} & 588\,G (46\%) & 381\,G (30\%) & 281\,G (22\%) & 223\,G (17\%) & 184\,G (14\%) \\
 & \multirow{2}{*}{Large} & WER (\%) & \textbf{4.31} & 4.63 & 16.20 & 72.34 & 105.4 & 109.3 \\
 & & GFLOPs & \textbf{2.6k\,G (100\%)} & 1.2k\,G (47\%) & 783\,G (30\%) & 579\,G (22\%) & 460\,G (18\%) & 381\,G (15\%) \\
\midrule
\multirow{10}{*}{ESD} & \multirow{2}{*}{Tiny} & WER (\%) & \textbf{14.16} & 35.87 & 163.8 & 228.8 & 189.1 & 200.8 \\
 & & GFLOPs & \textbf{39\,G (100\%)} & 16\,G (41\%) & 9.9\,G (26\%) & 7.2\,G (19\%) & 5.6\,G (14\%) & 4.6\,G (12\%) \\
 & \multirow{2}{*}{Base} & WER (\%) & \textbf{11.53} & 18.98 & 61.30 & 120.5 & 142.1 & 150.4 \\
 & & GFLOPs & \textbf{94\,G (100\%)} & 40\,G (43\%) & 25\,G (27\%) & 18\,G (20\%) & 14\,G (15\%) & 12\,G (13\%) \\
 & \multirow{2}{*}{Small} & WER (\%) & \textbf{9.39} & 13.20 & 38.48 & 82.77 & 102.5 & 108.9 \\
 & & GFLOPs & \textbf{381\,G (100\%)} & 170\,G (45\%) & 109\,G (29\%) & 80\,G (21\%) & 63\,G (17\%) & 52\,G (14\%) \\
 & \multirow{2}{*}{Medium} & WER (\%) & \textbf{8.65} & 11.22 & 26.50 & 66.67 & 91.46 & 97.93 \\
 & & GFLOPs & \textbf{1.3k\,G (100\%)} & 586\,G (46\%) & 379\,G (30\%) & 279\,G (22\%) & 221\,G (17\%) & 183\,G (14\%) \\
 & \multirow{2}{*}{Large} & WER (\%) & \textbf{9.63} & 10.16 & 22.00 & 63.18 & 96.61 & 140.9 \\
 & & GFLOPs & \textbf{2.6k\,G (100\%)} & 1.2k\,G (46\%) & 778\,G (30\%) & 577\,G (22\%) & 458\,G (18\%) & 380\,G (15\%) \\
\midrule
\multirow{10}{*}{Common Voice} & \multirow{2}{*}{Tiny} & WER (\%) & \textbf{35.85} & 88.69 & 234.7 & 260.1 & 224.0 & 191.7 \\
 & & GFLOPs & \textbf{39\,G (100\%)} & 16\,G (41\%) & 9.9\,G (26\%) & 7.2\,G (19\%) & 5.6\,G (14\%) & 4.6\,G (12\%) \\
 & \multirow{2}{*}{Base} & WER (\%) & \textbf{28.17} & 44.43 & 101.0 & 127.7 & 147.5 & 148.3 \\
 & & GFLOPs & \textbf{94\,G (100\%)} & 40\,G (43\%) & 25\,G (27\%) & 18\,G (20\%) & 14\,G (15\%) & 12\,G (13\%) \\
 & \multirow{2}{*}{Small} & WER (\%) & \textbf{17.93} & 29.10 & 66.06 & 104.5 & 111.1 & 115.4 \\
 & & GFLOPs & \textbf{381\,G (100\%)} & 170\,G (45\%) & 109\,G (29\%) & 80\,G (21\%) & 63\,G (17\%) & 52\,G (14\%) \\
 & \multirow{2}{*}{Medium} & WER (\%) & \textbf{14.26} & 20.84 & 48.53 & 85.07 & 97.73 & 99.49 \\
 & & GFLOPs & \textbf{1.3k\,G (100\%)} & 586\,G (46\%) & 379\,G (30\%) & 279\,G (22\%) & 221\,G (17\%) & 183\,G (14\%) \\
 & \multirow{2}{*}{Large} & WER (\%) & \textbf{11.66} & 17.56 & 44.91 & 83.98 & 102.7 & 115.3 \\
 & & GFLOPs & \textbf{2.6k\,G (100\%)} & 1.2k\,G (46\%) & 778\,G (30\%) & 576\,G (22\%) & 457\,G (18\%) & 379\,G (15\%) \\
\bottomrule
\end{tabular}
}
\caption{Input-side stride-$k$ subsampling: WER (\%), total GFLOPs with remaining ratio (\% of $k=1$), and total inference latency (ms) for all datasets, model scales, and stride values ($k = 1$--$6$). $k=1$ (\textbf{bold}) is the unmodified baseline.}
\label{tab:wer_input_full_2}
\end{table*}


\begin{table*}[t]
\centering
\small
\resizebox{\textwidth}{!}{%
\begin{tabular}{lllrrrrrr}
\toprule
\textbf{Dataset} & \textbf{Model} & \textbf{Metric} & $k=1$ & $k=2$ & $k=3$ & $k=4$ & $k=5$ & $k=6$ \\
\midrule
\multirow{10}{*}{LibriTTS} & \multirow{2}{*}{Tiny} & WER (\%) & \textbf{8.86} & 9.42 & 11.50 & 16.57 & 37.70 & 70.26 \\
 & & GFLOPs & \textbf{39\,G (100\%)} & 37\,G (95\%) & 36\,G (94\%) & 36\,G (93\%) & 36\,G (92\%) & 36\,G (92\%) \\
 & \multirow{2}{*}{Base} & WER (\%) & \textbf{6.87} & 6.76 & 7.11 & 8.22 & 15.15 & 47.60 \\
 & & GFLOPs & \textbf{94\,G (100\%)} & 89\,G (95\%) & 88\,G (93\%) & 87\,G (92\%) & 86\,G (92\%) & 86\,G (91\%) \\
 & \multirow{2}{*}{Small} & WER (\%) & \textbf{5.02} & 5.13 & 5.54 & 6.35 & 10.05 & 20.82 \\
 & & GFLOPs & \textbf{382\,G (100\%)} & 360\,G (94\%) & 353\,G (92\%) & 350\,G (91\%) & 347\,G (91\%) & 346\,G (90\%) \\
 & \multirow{2}{*}{Medium} & WER (\%) & \textbf{4.61} & 4.70 & 4.94 & 5.54 & 6.81 & 11.04 \\
 & & GFLOPs & \textbf{1.3k\,G (100\%)} & 1.2k\,G (94\%) & 1.2k\,G (92\%) & 1.2k\,G (91\%) & 1.2k\,G (90\%) & 1.2k\,G (90\%) \\
 & \multirow{2}{*}{Large} & WER (\%) & \textbf{4.31} & 4.30 & 4.69 & 5.47 & 10.66 & 29.56 \\
 & & GFLOPs & \textbf{2.6k\,G (100\%)} & 2.4k\,G (94\%) & 2.4k\,G (92\%) & 2.3k\,G (91\%) & 2.3k\,G (90\%) & 2.3k\,G (90\%) \\
\midrule
\multirow{10}{*}{ESD} & \multirow{2}{*}{Tiny} & WER (\%) & \textbf{14.16} & 14.66 & 15.72 & 17.71 & 25.94 & 43.51 \\
 & & GFLOPs & \textbf{39\,G (100\%)} & 37\,G (95\%) & 36\,G (94\%) & 36\,G (93\%) & 36\,G (92\%) & 36\,G (92\%) \\
 & \multirow{2}{*}{Base} & WER (\%) & \textbf{11.53} & 11.78 & 11.83 & 12.61 & 16.12 & 25.83 \\
 & & GFLOPs & \textbf{94\,G (100\%)} & 89\,G (95\%) & 88\,G (93\%) & 87\,G (92\%) & 86\,G (92\%) & 86\,G (91\%) \\
 & \multirow{2}{*}{Small} & WER (\%) & \textbf{9.39} & 9.53 & 9.75 & 10.45 & 11.59 & 16.75 \\
 & & GFLOPs & \textbf{381\,G (100\%)} & 360\,G (94\%) & 352\,G (92\%) & 349\,G (92\%) & 347\,G (91\%) & 345\,G (91\%) \\
 & \multirow{2}{*}{Medium} & WER (\%) & \textbf{8.65} & 8.81 & 9.05 & 9.77 & 10.58 & 13.00 \\
 & & GFLOPs & \textbf{1.3k\,G (100\%)} & 1.2k\,G (94\%) & 1.2k\,G (92\%) & 1.2k\,G (91\%) & 1.2k\,G (90\%) & 1.2k\,G (90\%) \\
 & \multirow{2}{*}{Large} & WER (\%) & \textbf{9.63} & 9.58 & 10.11 & 9.85 & 11.33 & 24.26 \\
 & & GFLOPs & \textbf{2.6k\,G (100\%)} & 2.4k\,G (94\%) & 2.4k\,G (92\%) & 2.3k\,G (91\%) & 2.3k\,G (90\%) & 2.3k\,G (90\%) \\
\midrule
\multirow{10}{*}{Common Voice} & \multirow{2}{*}{Tiny} & WER (\%) & \textbf{35.85} & 39.82 & 44.51 & 48.95 & 81.38 & 110.0 \\
 & & GFLOPs & \textbf{39\,G (100\%)} & 37\,G (95\%) & 36\,G (94\%) & 36\,G (93\%) & 36\,G (93\%) & 36\,G (92\%) \\
 & \multirow{2}{*}{Base} & WER (\%) & \textbf{28.17} & 28.48 & 31.35 & 32.34 & 38.43 & 52.19 \\
 & & GFLOPs & \textbf{94\,G (100\%)} & 89\,G (95\%) & 88\,G (93\%) & 87\,G (92\%) & 86\,G (92\%) & 86\,G (91\%) \\
 & \multirow{2}{*}{Small} & WER (\%) & \textbf{17.93} & 19.25 & 20.36 & 22.20 & 24.07 & 30.74 \\
 & & GFLOPs & \textbf{381\,G (100\%)} & 360\,G (94\%) & 352\,G (92\%) & 349\,G (92\%) & 347\,G (91\%) & 345\,G (91\%) \\
 & \multirow{2}{*}{Medium} & WER (\%) & \textbf{14.26} & 14.98 & 16.56 & 18.67 & 20.11 & 25.05 \\
 & & GFLOPs & \textbf{1.3k\,G (100\%)} & 1.2k\,G (94\%) & 1.2k\,G (92\%) & 1.2k\,G (91\%) & 1.2k\,G (90\%) & 1.2k\,G (90\%) \\
 & \multirow{2}{*}{Large} & WER (\%) & \textbf{11.66} & 12.70 & 14.36 & 16.26 & 21.03 & 30.95 \\
 & & GFLOPs & \textbf{2.6k\,G (100\%)} & 2.4k\,G (94\%) & 2.4k\,G (92\%) & 2.3k\,G (91\%) & 2.3k\,G (90\%) & 2.3k\,G (90\%) \\
\bottomrule
\end{tabular}
}
\caption{Output-side stride-$k$ subsampling: WER (\%), total GFLOPs with remaining ratio (\% of $k=1$), and total inference latency (ms) for all datasets, model scales, and stride values ($k = 1$--$6$). $k=1$ (\textbf{bold}) reproduces the unmodified baseline.}
\label{tab:wer_output_full_3}
\end{table*}

\begin{table*}[t]
\centering
\small
\setlength{\tabcolsep}{3.5pt}
\begin{tabular}{llrrrrr}
\toprule
Dataset & Model & Enc. FLOPs & Total FLOPs & Enc. ms & Total ms & WER \\
\midrule
\multirow{5}{*}{LibriTTS}
& Tiny & 14.07 (-20.99) & 15.08 (-23.83) & 2.08 (-0.37) & 116.57 (+8.58) & 28.48 (+19.62) \\
& Base & 35.22 (-49.05) & 37.87 (-56.50) & 3.04 (-0.42) & 147.19 (+13.90) & 14.13 (+7.25) \\
& Small & 148.14 (-189.61) & 159.89 (-222.58) & 5.71 (-0.41) & 233.63 (+25.48) & 9.22 (+4.20) \\
& Medium & 508.28 (-618.87) & 549.74 (-735.10) & 10.30 (-0.56) & 350.17 (+20.66) & 7.01 (+2.40) \\
& Large-v3 & 1035.88 (-1220.20) & 1121.84 (-1461.11) & 13.19 (-10.71) & 427.23 (-88.26) & 5.38 (+1.07) \\
\midrule
\multirow{5}{*}{ESD}
& Tiny & 14.07 (-20.99) & 15.02 (-23.72) & 2.30 (-0.80) & 66.05 (+1.33) & 41.18 (+27.02) \\
& Base & 35.22 (-49.05) & 37.71 (-56.29) & 3.14 (-1.70) & 68.74 (-10.17) & 20.36 (+8.83) \\
& Small & 148.14 (-189.61) & 159.26 (-221.96) & 5.60 (-2.97) & 94.38 (-21.07) & 14.23 (+4.84) \\
& Medium & 508.28 (-618.87) & 547.68 (-733.44) & 11.23 (-4.41) & 169.25 (-12.80) & 12.16 (+3.50) \\
& Large-v3 & 1035.88 (-1220.20) & 1117.78 (-1458.34) & 14.80 (-5.20) & 198.99 (-16.74) & 10.64 (+1.01) \\
\midrule
\multirow{5}{*}{Common Voice}
& Tiny & 14.07 (-20.99) & 15.06 (-23.68) & 2.45 (-0.52) & 143.77 (+67.02) & 113.70 (+77.85) \\
& Base & 35.22 (-49.05) & 37.75 (-56.25) & 3.36 (-0.69) & 116.12 (+29.61) & 47.65 (+19.48) \\
& Small & 148.14 (-189.61) & 159.39 (-221.75) & 5.88 (-1.65) & 149.41 (+5.00) & 31.34 (+13.40) \\
& Medium & 508.28 (-618.87) & 548.09 (-732.78) & 11.56 (-2.73) & 235.76 (-11.48) & 23.82 (+9.56) \\
& Large-v3 & 1035.88 (-1220.20) & 1118.86 (-1456.80) & 14.96 (-6.34) & 295.40 (-20.94) & 21.56 (+9.90) \\
\bottomrule
\end{tabular}
\caption{Compound subsampling results for $k_{\mathrm{enc\_in}}=k_{\mathrm{enc\_out}}=2$. Each cell reports the Both value with its delta from the EncInput baseline at $k=1$ in parentheses, computed as Both minus baseline. Total FLOPs are Enc. FLOPs + decoder-cross-attention FLOPs. WER is reported in percent.}
\label{tab:compound_full}
\end{table*}

\begin{table*}[t]
\centering
\begin{tabular}{lllrrrrr}
\toprule
\textbf{Model} & \textbf{Layer} & & $d=1$ & $d=2$ & $d=3$ & $d=4$ & $d=5$ \\
\midrule
\multirow{4}{*}{Whisper-Tiny} & \multirow{2}{*}{Conv Stem} & CKA & 0.9640 & 0.9037 & 0.8511 & 0.8117 & 0.7874 \\
 & & $\Delta$ & \textit{+0.0000} & \textit{-0.0603} & \textit{-0.1129} & \textit{-0.1523} & \textit{-0.1766} \\
\cmidrule(l){2-8}
 & \multirow{2}{*}{Enc Out} & CKA & 0.8808 & 0.7755 & 0.7377 & 0.7359 & 0.7386 \\
 & & $\Delta$ & \textit{+0.0000} & \textit{-0.1053} & \textit{-0.1432} & \textit{-0.1449} & \textit{-0.1422} \\
\midrule
\multirow{4}{*}{Whisper-Base} & \multirow{2}{*}{Conv Stem} & CKA & 0.9641 & 0.9016 & 0.8456 & 0.8029 & 0.7768 \\
 & & $\Delta$ & \textit{+0.0000} & \textit{-0.0624} & \textit{-0.1185} & \textit{-0.1612} & \textit{-0.1873} \\
\cmidrule(l){2-8}
 & \multirow{2}{*}{Enc Out} & CKA & 0.9030 & 0.8067 & 0.7628 & 0.7487 & 0.7528 \\
 & & $\Delta$ & \textit{+0.0000} & \textit{-0.0963} & \textit{-0.1402} & \textit{-0.1543} & \textit{-0.1501} \\
\midrule
\multirow{4}{*}{Whisper-Small} & \multirow{2}{*}{Conv Stem} & CKA & 0.9608 & 0.8945 & 0.8357 & 0.7916 & 0.7649 \\
 & & $\Delta$ & \textit{+0.0000} & \textit{-0.0663} & \textit{-0.1251} & \textit{-0.1692} & \textit{-0.1959} \\
\cmidrule(l){2-8}
 & \multirow{2}{*}{Enc Out} & CKA & 0.8277 & 0.7465 & 0.7539 & 0.7431 & 0.7666 \\
 & & $\Delta$ & \textit{+0.0000} & \textit{-0.0811} & \textit{-0.0737} & \textit{-0.0846} & \textit{-0.0610} \\
\midrule
\multirow{4}{*}{Whisper-Medium} & \multirow{2}{*}{Conv Stem} & CKA & 0.9619 & 0.9023 & 0.8516 & 0.8130 & 0.7883 \\
 & & $\Delta$ & \textit{+0.0000} & \textit{-0.0596} & \textit{-0.1103} & \textit{-0.1489} & \textit{-0.1736} \\
\cmidrule(l){2-8}
 & \multirow{2}{*}{Enc Out} & CKA & 0.7867 & 0.6622 & 0.6482 & 0.6505 & 0.6529 \\
 & & $\Delta$ & \textit{+0.0000} & \textit{-0.1245} & \textit{-0.1385} & \textit{-0.1362} & \textit{-0.1338} \\
\midrule
\multirow{4}{*}{Whisper-Large} & \multirow{2}{*}{Conv Stem} & CKA & 0.9795 & 0.9500 & 0.9250 & 0.9052 & 0.8926 \\
 & & $\Delta$ & \textit{+0.0000} & \textit{-0.0295} & \textit{-0.0545} & \textit{-0.0743} & \textit{-0.0869} \\
\cmidrule(l){2-8}
 & \multirow{2}{*}{Enc Out} & CKA & 0.8597 & 0.7629 & 0.7242 & 0.7133 & 0.7098 \\
 & & $\Delta$ & \textit{+0.0000} & \textit{-0.0968} & \textit{-0.1355} & \textit{-0.1463} & \textit{-0.1499} \\
\bottomrule
\end{tabular}
\caption{Linear CKA and cumulative drop $\Delta(d)=\text{CKA}(d)-\text{CKA}(1)$ for Conv Stem and Encoder outputs at frame distances $d=1$--$5$. Conv Stem similarity decreases monotonically, whereas Encoder output similarity converges after $d=2$--$3$. Computed on LibriTTS}
\label{tab:cka_drop}
\end{table*}


\begin{table*}[t]
\centering
\small
\renewcommand{\arraystretch}{1}
\setlength{\cmidrulewidth}{0.5pt}

\begin{tabular}{
l
l
r
r
r
r
r
r
}
\toprule
\textbf{Category} & \textbf{Method} &
\textbf{Acc.} & \textbf{$\Delta$A} & \textbf{FLOPs} &
\textbf{$\Delta$F} & \textbf{Lat.} & \textbf{$\Delta$L} \\
& & \textbf{(\%)} & \textbf{(pt)} & \textbf{(G)} &
\textbf{(\%)} & \textbf{(ms)} & \textbf{(\%)} \\
\midrule

\multirow{10}{*}{Reasoning} & baseline $k{=}1$ & 77.19 & -- & 5662.56 & -- & 103.00 & -- \\
\cmidrule(lr){2-8}
& Enc $k{=}2$ & 76.28 & -0.91 & 3419.02 & -39.62 & 84.45 & -18.01 \\
\cmidrule(lr){2-8}
& SP $r{=}0.2$ & 75.58 & -1.61 & 5229.53 & -7.65 & 85.07 & -17.41 \\
& SP $r{=}0.4$ & 71.45 & -5.74 & 4825.53 & -14.78 & 80.39 & -21.95 \\
& SP $r{=}0.6$ & 67.57 & -9.62 & 4423.59 & -21.88 & 76.01 & -26.20 \\
& SP $r{=}0.8$ & 60.66 & -16.53 & 4018.48 & -29.03 & 72.86 & -29.27 \\
\cmidrule(lr){2-8}
& Enc+SP $r{=}0.2$ & 68.06 & -9.13 & 3202.44 & -43.45 & 74.36 & -27.80 \\
& Enc+SP $r{=}0.4$ & 64.38 & -12.81 & 3002.08 & -46.98 & 72.96 & -29.16 \\
& Enc+SP $r{=}0.6$ & 61.65 & -15.54 & 2801.13 & -50.53 & 71.77 & -30.32 \\
& Enc+SP $r{=}0.8$ & 58.10 & -19.09 & 2593.18 & -54.20 & 70.19 & -31.86 \\
\midrule

\multirow{10}{*}{Perception} & baseline $k{=}1$ & 42.67 & -- & 5924.75 & -- & 106.90 & -- \\
\cmidrule(lr){2-8}
& Enc $k{=}2$ & 42.36 & -0.31 & 3396.50 & -42.67 & 86.12 & -19.44 \\
\cmidrule(lr){2-8}
& SP $r{=}0.2$ & 41.82 & -0.85 & 5377.61 & -9.23 & 87.97 & -17.71 \\
& SP $r{=}0.4$ & 41.71 & -0.96 & 4860.47 & -17.96 & 82.61 & -22.72 \\
& SP $r{=}0.6$ & 40.31 & -2.36 & 4341.18 & -26.73 & 77.09 & -27.88 \\
& SP $r{=}0.8$ & 37.67 & -5.00 & 3824.41 & -35.45 & 73.92 & -30.85 \\
\cmidrule(lr){2-8}
& Enc+SP $r{=}0.2$ & 40.97 & -1.70 & 3120.62 & -47.33 & 77.70 & -27.31 \\
& Enc+SP $r{=}0.4$ & 39.42 & -3.25 & 2863.55 & -51.67 & 75.96 & -28.94 \\
& Enc+SP $r{=}0.6$ & 36.67 & -6.00 & 2606.17 & -56.01 & 74.64 & -30.18 \\
& Enc+SP $r{=}0.8$ & 35.23 & -7.44 & 2351.79 & -60.31 & 73.98 & -30.79 \\

\bottomrule
\end{tabular}
\caption{Category-level results on MMSU for \textbf{Audioflamingo3}. We report accuracy, FLOPs, and latency for each category. Enc $k{=}2$ denotes encoder-side stride-2 subsampling. SP denotes SpeechPrune, where $r$ indicates the pruning ratio (e.g., $r{=}0.2$ removes 20\% of speech tokens). Enc+SP denotes the combined setting. All relative changes are computed with respect to the corresponding baseline within the same category.}
\label{tab:mmsu_audioflamingo3}
\end{table*}

\begin{table*}[t]
\centering
\small
\renewcommand{\arraystretch}{1}
\setlength{\cmidrulewidth}{0.5pt}
\begin{tabular}{
l
l
r
r
r
r
r
r
}
\toprule
\textbf{Category} & \textbf{Method} &
\textbf{Acc.} & \textbf{$\Delta$A} & \textbf{FLOPs} &
\textbf{$\Delta$F} & \textbf{Lat.} & \textbf{$\Delta$L} \\
& & \textbf{(\%)} & \textbf{(pt)} & \textbf{(G)} &
\textbf{(\%)} & \textbf{(ms)} & \textbf{(\%)} \\
\midrule
\multirow{10}{*}{Reasoning} & baseline $k{=}1$ & 58.47 & -- & 7594.48 & -- & 179.06 & -- \\
\cmidrule(lr){2-8}
& Enc $k{=}2$ & 56.40 & -2.07 & 4373.39 & -42.41 & 126.18 & -29.53 \\
\cmidrule(lr){2-8}
& SP $r{=}0.2$ & 56.77 & -1.70 & 6797.79 & -10.49 & 132.55 & -25.97 \\
& SP $r{=}0.4$ & 57.11 & -1.36 & 5999.92 & -21.00 & 127.57 & -28.75 \\
& SP $r{=}0.6$ & 54.75 & -3.72 & 5210.78 & -31.39 & 129.88 & -27.46 \\
& SP $r{=}0.8$ & 49.67 & -8.80 & 4409.54 & -41.94 & 127.29 & -28.91 \\
\cmidrule(lr){2-8}
& Enc+SP $r{=}0.2$ & 56.03 & -2.44 & 3979.94 & -47.59 & 83.71 & -53.25 \\
& Enc+SP $r{=}0.4$ & 55.21 & -3.26 & 3584.12 & -52.81 & 81.15 & -54.68 \\
& Enc+SP $r{=}0.6$ & 51.69 & -6.78 & 3193.98 & -57.94 & 85.09 & -52.48 \\
& Enc+SP $r{=}0.8$ & 49.46 & -9.01 & 2792.94 & -63.22 & 89.49 & -50.02 \\
\midrule
\multirow{10}{*}{Perception} & baseline $k{=}1$ & 29.96 & -- & 7342.39 & -- & 215.46 & -- \\
\cmidrule(lr){2-8}
& Enc $k{=}2$ & 28.45 & -1.51 & 4107.21 & -44.06 & 140.40 & -34.84 \\
\cmidrule(lr){2-8}
& SP $r{=}0.2$ & 28.45 & -1.51 & 6553.31 & -10.75 & 178.10 & -17.34 \\
& SP $r{=}0.4$ & 28.84 & -1.12 & 5750.38 & -21.68 & 169.34 & -21.40 \\
& SP $r{=}0.6$ & 28.95 & -1.01 & 4967.55 & -32.34 & 190.83 & -11.43 \\
& SP $r{=}0.8$ & 27.95 & -2.01 & 4162.83 & -43.30 & 195.42 & -9.30 \\
\cmidrule(lr){2-8}
& Enc+SP $r{=}0.2$ & 28.45 & -1.51 & 3717.57 & -49.37 & 107.78 & -49.98 \\
& Enc+SP $r{=}0.4$ & 27.91 & -2.05 & 3320.87 & -54.77 & 108.25 & -49.76 \\
& Enc+SP $r{=}0.6$ & 28.06 & -1.90 & 2941.09 & -59.94 & 139.85 & -35.09 \\
& Enc+SP $r{=}0.8$ & 27.13 & -2.83 & 2544.03 & -65.35 & 159.05 & -26.18 \\
\bottomrule
\end{tabular}
\caption{Category-level results on MMSU for \textbf{LLaMA-Omni2}. We report accuracy, FLOPs, and latency for each category. Enc $k{=}2$ denotes encoder-side stride-2 subsampling. SP denotes SpeechPrune, where $r$ indicates the pruning ratio (e.g., $r{=}0.2$ removes 20\% of speech tokens). Enc+SP denotes the combined setting. All relative changes are computed with respect to the corresponding baseline within the same category.}
\label{tab:mmsu_llamaomni2}
\end{table*}

\begin{table*}[t]
\centering
\small
\renewcommand{\arraystretch}{1}
\setlength{\cmidrulewidth}{0.5pt}
\begin{tabular}{
l
l
r
r
r
r
r
r
}
\toprule
\textbf{Category} & \textbf{Method} &
\textbf{Acc.} & \textbf{$\Delta$A} & \textbf{FLOPs} &
\textbf{$\Delta$F} & \textbf{Lat.} & \textbf{$\Delta$L} \\
& & \textbf{(\%)} & \textbf{(pt)} & \textbf{(G)} &
\textbf{(\%)} & \textbf{(ms)} & \textbf{(\%)} \\
\midrule
\multirow{10}{*}{Reasoning} & baseline $k{=}1$ & 69.71 & -- & 5736.51 & -- & 108.43 & -- \\
\cmidrule(lr){2-8}
& Enc $k{=}2$ & 68.39 & -1.32 & 3503.23 & -38.93 & 92.75 & -14.46 \\
\cmidrule(lr){2-8}
& SP $r{=}0.2$ & 68.47 & -1.24 & 5318.06 & -7.29 & 89.12 & -17.81 \\
& SP $r{=}0.4$ & 63.93 & -5.78 & 4916.16 & -14.30 & 83.98 & -22.55 \\
& SP $r{=}0.6$ & 57.44 & -12.27 & 4518.10 & -21.24 & 84.77 & -21.82 \\
& SP $r{=}0.8$ & 50.74 & -18.97 & 4115.58 & -28.26 & 83.71 & -22.79 \\
\cmidrule(lr){2-8}
& Enc+SP $r{=}0.2$ & 66.66 & -3.05 & 3296.68 & -42.53 & 81.13 & -25.17 \\
& Enc+SP $r{=}0.4$ & 62.36 & -7.35 & 3099.17 & -45.97 & 82.77 & -23.67 \\
& Enc+SP $r{=}0.6$ & 55.37 & -14.34 & 2901.52 & -49.42 & 86.48 & -20.25 \\
& Enc+SP $r{=}0.8$ & 48.64 & -21.07 & 2697.09 & -52.98 & 90.21 & -16.80 \\
\midrule
\multirow{10}{*}{Perception} & baseline $k{=}1$ & 41.12 & -- & 5994.74 & -- & 110.48 & -- \\
\cmidrule(lr){2-8}
& Enc $k{=}2$ & 40.35 & -0.77 & 3477.79 & -41.99 & 91.58 & -17.11 \\
\cmidrule(lr){2-8}
& SP $r{=}0.2$ & 40.04 & -1.08 & 5461.46 & -8.90 & 86.00 & -22.16 \\
& SP $r{=}0.4$ & 38.33 & -2.79 & 4948.49 & -17.45 & 81.67 & -26.08 \\
& SP $r{=}0.6$ & 36.82 & -4.30 & 4431.98 & -26.07 & 76.75 & -30.53 \\
& SP $r{=}0.8$ & 32.95 & -8.17 & 3918.70 & -34.63 & 74.05 & -32.98 \\
\cmidrule(lr){2-8}
& Enc+SP $r{=}0.2$ & 38.91 & -2.21 & 3210.58 & -46.44 & 72.97 & -33.96 \\
& Enc+SP $r{=}0.4$ & 36.82 & -4.30 & 2955.78 & -50.69 & 71.65 & -35.15 \\
& Enc+SP $r{=}0.6$ & 35.27 & -5.85 & 2700.18 & -54.96 & 70.47 & -36.22 \\
& Enc+SP $r{=}0.8$ & 32.68 & -8.44 & 2447.55 & -59.17 & 70.44 & -36.24 \\
\bottomrule
\end{tabular}
\caption{Category-level results on MMSU for \textbf{Qwen2-Audio}. We report accuracy, FLOPs, and latency for each category. Enc $k{=}2$ denotes encoder-side stride-2 subsampling. SP denotes SpeechPrune, where $r$ indicates the pruning ratio (e.g., $r{=}0.2$ removes 20\% of speech tokens). Enc+SP denotes the combined setting. All relative changes are computed with respect to the corresponding baseline within the same category.}
\label{tab:mmsu_qwen2audio}
\end{table*}

\begin{table*}[t]
\centering
\small
\renewcommand{\arraystretch}{1}
\setlength{\cmidrulewidth}{0.5pt}
\begin{tabular}{
l
l
r
r
r
r
r
r
}
\toprule
\textbf{Category} & \textbf{Method} &
\textbf{Acc.} & \textbf{$\Delta$A} & \textbf{FLOPs} &
\textbf{$\Delta$F} & \textbf{Lat.} & \textbf{$\Delta$L} \\
& & \textbf{(\%)} & \textbf{(pt)} & \textbf{(G)} &
\textbf{(\%)} & \textbf{(ms)} & \textbf{(\%)} \\
\midrule
\multirow{10}{*}{Sound} & baseline $k{=}1$ & 73.57 & -- & 6320.28 & -- & 190.04 & -- \\
\cmidrule(lr){2-8}
& Enc $k{=}2$ & 73.27 & -0.30 & 3617.51 & -42.76 & 166.08 & -12.61 \\
\cmidrule(lr){2-8}
& SP $r{=}0.2$ & 73.87 & 0.30 & 5712.18 & -9.62 & 186.04 & -2.10 \\
& SP $r{=}0.4$ & 73.57 & 0.00 & 5123.17 & -18.94 & 174.19 & -8.34 \\
& SP $r{=}0.6$ & 73.27 & -0.30 & 4538.63 & -28.19 & 168.49 & -11.34 \\
& SP $r{=}0.8$ & 72.07 & -1.50 & 3949.32 & -37.51 & 164.29 & -13.55 \\
\cmidrule(lr){2-8}
& Enc+SP $r{=}0.2$ & 72.97 & -0.60 & 3319.03 & -47.49 & 173.39 & -8.76 \\
& Enc+SP $r{=}0.4$ & 73.57 & 0.00 & 3028.62 & -52.08 & 172.01 & -9.49 \\
& Enc+SP $r{=}0.6$ & 72.07 & -1.50 & 2740.01 & -56.65 & 169.52 & -10.80 \\
& Enc+SP $r{=}0.8$ & 70.87 & -2.70 & 2438.29 & -61.42 & 167.57 & -11.82 \\
\midrule
\multirow{10}{*}{Speech} & baseline $k{=}1$ & 58.56 & -- & 8298.17 & -- & 199.73 & -- \\
\cmidrule(lr){2-8}
& Enc $k{=}2$ & 57.36 & -1.20 & 4601.48 & -44.55 & 175.57 & -12.10 \\
\cmidrule(lr){2-8}
& SP $r{=}0.2$ & 57.36 & -1.20 & 7285.25 & -12.21 & 195.30 & -2.22 \\
& SP $r{=}0.4$ & 55.56 & -3.00 & 6297.97 & -24.10 & 185.99 & -6.88 \\
& SP $r{=}0.6$ & 54.05 & -4.51 & 5318.16 & -35.91 & 177.70 & -11.03 \\
& SP $r{=}0.8$ & 48.65 & -9.91 & 4351.36 & -47.56 & 169.85 & -14.96 \\
\cmidrule(lr){2-8}
& Enc+SP $r{=}0.2$ & 56.76 & -1.80 & 4099.12 & -50.60 & 170.16 & -14.80 \\
& Enc+SP $r{=}0.4$ & 56.16 & -2.40 & 3613.18 & -56.46 & 167.28 & -16.25 \\
& Enc+SP $r{=}0.6$ & 54.05 & -4.51 & 3130.75 & -62.27 & 163.78 & -18.00 \\
& Enc+SP $r{=}0.8$ & 46.25 & -12.31 & 2658.61 & -67.96 & 162.53 & -18.63 \\
\midrule
\multirow{10}{*}{Music} & baseline $k{=}1$ & 79.34 & -- & 9354.09 & -- & 233.91 & -- \\
\cmidrule(lr){2-8}
& Enc $k{=}2$ & 78.14 & -1.20 & 5202.11 & -44.39 & 201.50 & -13.86 \\
\cmidrule(lr){2-8}
& SP $r{=}0.2$ & 78.44 & -0.90 & 8165.85 & -12.70 & 226.35 & -3.23 \\
& SP $r{=}0.4$ & 78.14 & -1.20 & 6997.61 & -25.19 & 212.83 & -9.01 \\
& SP $r{=}0.6$ & 76.35 & -2.99 & 5832.51 & -37.65 & 203.48 & -13.01 \\
& SP $r{=}0.8$ & 76.35 & -2.99 & 4648.42 & -50.31 & 195.48 & -16.43 \\
\cmidrule(lr){2-8}
& Enc+SP $r{=}0.2$ & 76.65 & -2.69 & 4611.91 & -50.70 & 194.58 & -16.81 \\
& Enc+SP $r{=}0.4$ & 77.54 & -1.80 & 4029.30 & -56.92 & 191.92 & -17.95 \\
& Enc+SP $r{=}0.6$ & 76.35 & -2.99 & 3444.30 & -63.18 & 187.16 & -19.99 \\
& Enc+SP $r{=}0.8$ & 76.35 & -2.99 & 2841.26 & -69.63 & 184.72 & -21.03 \\
\bottomrule
\end{tabular}
\caption{Category-level results on MMAU for \textbf{Audioflamingo3}. We report accuracy, FLOPs, and latency for each category. Enc $k{=}2$ denotes encoder-side stride-2 subsampling. SP denotes SpeechPrune, where $r$ indicates the pruning ratio (e.g., $r{=}0.2$ removes 20\% of speech tokens). Enc+SP denotes the combined setting. All relative changes are computed with respect to the corresponding baseline within the same category.}
\label{tab:mmau_audioflamingo3}
\end{table*}

\begin{table*}[t]
\centering
\small
\renewcommand{\arraystretch}{1}
\setlength{\cmidrulewidth}{0.5pt}
\begin{tabular}{
l
l
r
r
r
r
r
r
}
\toprule
\textbf{Category} & \textbf{Method} &
\textbf{Acc.} & \textbf{$\Delta$A} & \textbf{FLOPs} &
\textbf{$\Delta$F} & \textbf{Lat.} & \textbf{$\Delta$L} \\
& & \textbf{(\%)} & \textbf{(pt)} & \textbf{(G)} &
\textbf{(\%)} & \textbf{(ms)} & \textbf{(\%)} \\
\midrule
\multirow{10}{*}{Sound} & baseline $k{=}1$ & 52.55 & -- & 7331.06 & -- & 169.65 & -- \\
\cmidrule(lr){2-8}
& Enc $k{=}2$ & 48.05 & -4.50 & 4094.61 & -44.15 & 96.91 & -42.88 \\
\cmidrule(lr){2-8}
& SP $r{=}0.2$ & 48.35 & -4.20 & 6519.56 & -11.07 & 147.91 & -12.81 \\
& SP $r{=}0.4$ & 51.35 & -1.20 & 5728.46 & -21.86 & 158.14 & -6.78 \\
& SP $r{=}0.6$ & 51.35 & -1.20 & 4923.42 & -32.84 & 138.04 & -18.63 \\
& SP $r{=}0.8$ & 51.35 & -1.20 & 4121.66 & -43.78 & 143.88 & -15.19 \\
\cmidrule(lr){2-8}
& Enc+SP $r{=}0.2$ & 45.65 & -6.90 & 3695.95 & -49.59 & 93.46 & -44.91 \\
& Enc+SP $r{=}0.4$ & 51.95 & -0.60 & 3301.89 & -54.96 & 98.34 & -42.03 \\
& Enc+SP $r{=}0.6$ & 51.35 & -1.20 & 2905.21 & -60.37 & 97.91 & -42.29 \\
& Enc+SP $r{=}0.8$ & 52.25 & -0.30 & 2496.73 & -65.94 & 97.32 & -42.63 \\
\midrule
\multirow{10}{*}{Speech} & baseline $k{=}1$ & 48.95 & -- & 7382.42 & -- & 186.66 & -- \\
\cmidrule(lr){2-8}
& Enc $k{=}2$ & 46.55 & -2.40 & 4150.85 & -43.77 & 120.84 & -35.26 \\
\cmidrule(lr){2-8}
& SP $r{=}0.2$ & 44.44 & -4.51 & 6579.29 & -10.88 & 177.77 & -4.76 \\
& SP $r{=}0.4$ & 42.04 & -6.91 & 5789.22 & -21.58 & 186.73 & 0.04 \\
& SP $r{=}0.6$ & 42.64 & -6.31 & 4997.85 & -32.30 & 187.25 & 0.32 \\
& SP $r{=}0.8$ & 39.64 & -9.31 & 4192.79 & -43.21 & 181.80 & -2.60 \\
\cmidrule(lr){2-8}
& Enc+SP $r{=}0.2$ & 45.05 & -3.90 & 3757.19 & -49.11 & 122.26 & -34.50 \\
& Enc+SP $r{=}0.4$ & 45.05 & -3.90 & 3370.60 & -54.34 & 136.32 & -26.97 \\
& Enc+SP $r{=}0.6$ & 41.14 & -7.81 & 2984.20 & -59.58 & 149.77 & -19.76 \\
& Enc+SP $r{=}0.8$ & 37.54 & -11.41 & 2567.95 & -65.22 & 130.84 & -29.90 \\
\midrule
\multirow{10}{*}{Music} & baseline $k{=}1$ & 52.69 & -- & 7417.40 & -- & 174.36 & -- \\
\cmidrule(lr){2-8}
& Enc $k{=}2$ & 50.30 & -2.39 & 4196.86 & -43.42 & 126.39 & -27.51 \\
\cmidrule(lr){2-8}
& SP $r{=}0.2$ & 52.40 & -0.29 & 6614.35 & -10.83 & 166.52 & -4.50 \\
& SP $r{=}0.4$ & 51.50 & -1.19 & 5815.88 & -21.59 & 162.26 & -6.94 \\
& SP $r{=}0.6$ & 50.60 & -2.09 & 5025.23 & -32.25 & 164.74 & -5.52 \\
& SP $r{=}0.8$ & 51.80 & -0.89 & 4212.47 & -43.21 & 146.22 & -16.14 \\
\cmidrule(lr){2-8}
& Enc+SP $r{=}0.2$ & 51.20 & -1.49 & 3803.46 & -48.72 & 129.71 & -25.61 \\
& Enc+SP $r{=}0.4$ & 51.50 & -1.19 & 3398.87 & -54.18 & 114.03 & -34.60 \\
& Enc+SP $r{=}0.6$ & 51.50 & -1.19 & 2999.00 & -59.57 & 105.35 & -39.58 \\
& Enc+SP $r{=}0.8$ & 51.80 & -0.89 & 2593.66 & -65.03 & 105.48 & -39.50 \\
\bottomrule
\end{tabular}
\caption{Category-level results on MMAU for \textbf{LLaMA-Omni2}. We report accuracy, FLOPs, and latency for each category. Enc $k{=}2$ denotes encoder-side stride-2 subsampling. SP denotes SpeechPrune, where $r$ indicates the pruning ratio (e.g., $r{=}0.2$ removes 20\% of speech tokens). Enc+SP denotes the combined setting. All relative changes are computed with respect to the corresponding baseline within the same category.}
\label{tab:mmau_llamaomni2}
\end{table*}

\begin{table*}[t]
\centering
\small
\renewcommand{\arraystretch}{1}
\setlength{\cmidrulewidth}{0.5pt}
\begin{tabular}{
l
l
r
r
r
r
r
r
}
\toprule
\textbf{Category} & \textbf{Method} &
\textbf{Acc.} & \textbf{$\Delta$A} & \textbf{FLOPs} &
\textbf{$\Delta$F} & \textbf{Lat.} & \textbf{$\Delta$L} \\
& & \textbf{(\%)} & \textbf{(pt)} & \textbf{(G)} &
\textbf{(\%)} & \textbf{(ms)} & \textbf{(\%)} \\
\midrule
\multirow{10}{*}{Sound} & baseline $k{=}1$ & 59.76 & -- & 6369.72 & -- & 132.04 & -- \\
\cmidrule(lr){2-8}
& Enc $k{=}2$ & 57.06 & -2.70 & 3668.80 & -42.40 & 101.86 & -22.86 \\
\cmidrule(lr){2-8}
& SP $r{=}0.2$ & 59.76 & -0.00 & 5767.23 & -9.46 & 127.47 & -3.46 \\
& SP $r{=}0.4$ & 59.16 & -0.60 & 5181.65 & -18.65 & 118.39 & -10.34 \\
& SP $r{=}0.6$ & 56.16 & -3.60 & 4595.33 & -27.86 & 105.55 & -20.06 \\
& SP $r{=}0.8$ & 52.55 & -7.21 & 4010.33 & -37.04 & 104.76 & -20.66 \\
\cmidrule(lr){2-8}
& Enc+SP $r{=}0.2$ & 56.76 & -3.00 & 3373.63 & -47.04 & 94.99 & -28.06 \\
& Enc+SP $r{=}0.4$ & 55.56 & -4.20 & 3085.67 & -51.56 & 95.44 & -27.72 \\
& Enc+SP $r{=}0.6$ & 57.06 & -2.70 & 2797.44 & -56.08 & 93.65 & -29.07 \\
& Enc+SP $r{=}0.8$ & 51.05 & -8.71 & 2501.26 & -60.73 & 101.44 & -23.17 \\
\midrule
\multirow{10}{*}{Speech} & baseline $k{=}1$ & 53.75 & -- & 8303.95 & -- & 113.94 & -- \\
\cmidrule(lr){2-8}
& Enc $k{=}2$ & 53.15 & -0.60 & 4615.62 & -44.42 & 94.24 & -17.29 \\
\cmidrule(lr){2-8}
& SP $r{=}0.2$ & 52.25 & -1.50 & 7300.69 & -12.08 & 115.83 & 1.66 \\
& SP $r{=}0.4$ & 50.45 & -3.30 & 6315.84 & -23.94 & 109.96 & -3.49 \\
& SP $r{=}0.6$ & 45.05 & -8.70 & 5341.69 & -35.67 & 101.95 & -10.52 \\
& SP $r{=}0.8$ & 37.24 & -16.51 & 4382.80 & -47.22 & 95.79 & -15.93 \\
\cmidrule(lr){2-8}
& Enc+SP $r{=}0.2$ & 51.65 & -2.10 & 4120.97 & -50.37 & 93.01 & -18.37 \\
& Enc+SP $r{=}0.4$ & 45.35 & -8.40 & 3639.63 & -56.17 & 93.64 & -17.82 \\
& Enc+SP $r{=}0.6$ & 41.14 & -12.61 & 3164.31 & -61.89 & 95.45 & -16.23 \\
& Enc+SP $r{=}0.8$ & 32.73 & -21.02 & 2705.37 & -67.42 & 107.65 & -5.52 \\
\midrule
\multirow{10}{*}{Music} & baseline $k{=}1$ & 55.39 & -- & 9358.64 & -- & 132.06 & -- \\
\cmidrule(lr){2-8}
& Enc $k{=}2$ & 55.39 & 0.00 & 5212.75 & -44.30 & 101.72 & -22.97 \\
\cmidrule(lr){2-8}
& SP $r{=}0.2$ & 53.29 & -2.10 & 8183.63 & -12.56 & 135.59 & 2.67 \\
& SP $r{=}0.4$ & 52.99 & -2.40 & 7017.12 & -25.02 & 120.91 & -8.44 \\
& SP $r{=}0.6$ & 53.29 & -2.10 & 5853.33 & -37.46 & 108.11 & -18.14 \\
& SP $r{=}0.8$ & 52.40 & -2.99 & 4673.40 & -50.06 & 97.26 & -26.35 \\
\cmidrule(lr){2-8}
& Enc+SP $r{=}0.2$ & 54.49 & -0.90 & 4631.25 & -50.51 & 99.23 & -24.86 \\
& Enc+SP $r{=}0.4$ & 55.09 & -0.30 & 4049.74 & -56.73 & 94.22 & -28.65 \\
& Enc+SP $r{=}0.6$ & 52.99 & -2.40 & 3469.70 & -62.93 & 94.25 & -28.63 \\
& Enc+SP $r{=}0.8$ & 50.30 & -5.09 & 2869.43 & -69.34 & 92.05 & -30.30 \\
\bottomrule
\end{tabular}
\caption{Category-level results on MMAU for \textbf{Qwen2-Audio}. We report accuracy, FLOPs, and latency for each category. Enc $k{=}2$ denotes encoder-side stride-2 subsampling. SP denotes SpeechPrune, where $r$ indicates the pruning ratio (e.g., $r{=}0.2$ removes 20\% of speech tokens). Enc+SP denotes the combined setting. All relative changes are computed with respect to the corresponding baseline within the same category.}
\label{tab:mmau_qwenaudio}
\end{table*}

\begin{table*}[t]
\centering
\small
\setlength{\tabcolsep}{5pt}
\renewcommand{\arraystretch}{1.05}

\begin{tabular}{llrr}
\toprule
Dataset & Mode & WER (\%) & Total FLOPs (G) \\
\midrule

\multirow{3}{*}{LibriTTS}
& Distil & 4.94 & 2276.40 \\
& Distil + EncInput $k=2$ & 5.93 & 1046.19 \\
& Distil + compound $k=2$ & 8.31 & 1041.17 \\
\midrule

\multirow{3}{*}{ESD}
& Distil & 9.48 & 2275.97 \\
& Distil + EncInput $k=2$ & 10.69 & 1045.88 \\
& Distil + compound $k=2$ & 13.28 & 1040.93 \\
\midrule

\multirow{3}{*}{Common Voice}
& Distil & 14.51 & 2276.05 \\
& Distil + EncInput $k=2$ & 23.65 & 1045.94 \\
& Distil + compound $k=2$ & 31.07 & 1041.00 \\

\bottomrule
\end{tabular}

\caption{Distil-Whisper large-v3 ASR performance and total FLOPs under stride-$k$ sampling.}
\label{tab:adistilwhisper}
\end{table*}